\documentclass[12pt,preprint]{aastex}

\usepackage{amsmath}

\def\gsim{\;\rlap{\lower 2.5pt
 \hbox{$\sim$}}\raise 1.5pt\hbox{$>$}\;}
\def\lsim{\;\rlap{\lower 2.5pt
   \hbox{$\sim$}}\raise 1.5pt\hbox{$<$}\;}

\newcommand{\be}{\begin{equation}}
\newcommand{\beq}{\begin{equation}}
\newcommand{\ba}{\begin{eqnarray}}
\newcommand{\ee}{\end{equation}}
\newcommand{\eeq}{\end{equation}}
\newcommand{\ea}{\end{eqnarray}}

\usepackage{epsfig}

\begin{document}
%\twocolumn[%%% Begin front material
%\submitted{Submitted to ApJ}
\title{A low-redshift galaxy cluster X-ray Temperature Function incorporating {\sl Suzaku} data}

\author{Cien Shang}
\affil{Department of Physics, Columbia University, 550 West 120th
Street, New York, NY 10027, USA; cien@phys.columbia.edu}

\author{Caleb Scharf}
\affil{Columbia Astrophysics Laboratory, Columbia University, 550 West
120th Street, New York, NY 10027, USA; caleb@astro.columbia.edu}

\begin{abstract}
We present the first results of a survey of 14 low redshift galaxy clusters using {\sl Suzaku}. Although luminous ($L_x> 1\times 10^{43}$ erg s$^{-1}$ (0.1-2.4 keV)), these clusters have no prior pointed X-ray data. Together with 47 other systems they form a flux limited sample ($f_x >1.0 \times 10^{-11}$ erg s$^{-1}$cm$^{-2}$ (0.1-2.4 keV)) with $z\leq 0.1$ in the northern celestial hemisphere. Using this total sample we evaluate the local $L-T$ relationship and the local cluster temperature function. {\sl Suzaku} temperature measurements appear to be in accord with those of other missions. General agreement is found with other published estimates of the low redshift cluster temperature function, however the sample used here exhibits slightly lower space densities at gas temperatures below 4-5 keV. We find a corresponding deficit in the number of clusters with temperatures between approximately 4.5 and 5.5 keV. Although at low significance, a similar feature exists in previous low redshift cluster datasets. We suggest that low-redshift cluster samples, while crucial for calibrating precision cosmology measurements, must be used with caution due to their limited size and susceptibility to the effects of cosmic variance.

\end{abstract}
\keywords{cosmology: observations -- X-rays: galaxies: clusters -- intergalactic medium}
%\vspace{3\baselineskip}
%%% End front material

\section{Introduction}
\label{sec:introduction}

The nature and evolution of galaxy clusters offers an independent and complementary probe of
cosmology to those of the cosmic microwave background \citep{hinshaw08} or supernova (e.g. Permlmutter \& Schmidt 2003). Theoretical work (e.g. Haiman, Mohr \& Holder 2001) has shown that  large surveys of clusters to reasonably high redshift ($z\sim 1$) can not only provide these much needed constraints, but
can, through the use of information such as the spatial distribution of clusters, be made
``self-calibrating'' - {\em if} the evolution of cluster structure can be parameterized (Hu
2003, Majumdar \& Mohr 2003), and {\em if} the scatter between cluster mass and observables
(luminosity, temperature) is understood for clusters {\em at the survey detection
threshold}.

Future X-ray surveys or Sunyaev-Zel'dovich Effect (SZE) surveys, such as the South Pole Telescope (SPT),
and the Atacama Cosmology Telescope (ACT), will be designed to detect $\sim 10^4$ clusters
to high redshift within a limited sky area and will not sample very many local clusters, or
obtain very high physical resolution data. However, both of these characteristics are vital
in order to establish the zero-redshift baseline from which we can then evaluate the global
evolution of cluster scaling laws and structure, and calibrate possible systematic effects
induced by potentially complex astrophysics.

In this paper we describe the results of an effort to utilize existing X-ray cluster data, and new observations by the {\sl Suzaku} (JAXA/NASA)  mission to add to the existing body of work on the low redshift population of clusters (e.g. Henry 2000; Reiprich \& Bohringer 2002; Ikebe et al. 2002). Specifically, we present the results of a {\sl Suzaku} survey of 14 nearby, X-ray luminous, clusters that have otherwise {\em never} been observed in a direct, pointed X-ray observation with earlier, spectrally sensitive instruments. Our motivation is to increase the robustness of the local cluster temperature
function at {\em lower} masses, which are typical of the mass threshold of the majority of
likely future X-ray and SZE cluster surveys. The range of masses probed by this data combined with pre-existing
 data (a total, complete, sample of 61 objects) is approximately $8\times 10^{13}$---$10^{15}$ M$_{\odot}$ ($H_0=70$). We therefore bracket the expected low mass limit of future surveys, all of which are designed to probe to $\sim
2\times 10^{14}$ M$_{\odot}$ (locally for the X-ray, and across essentially all redshifts for
SPT/ACT). For modern cosmological tests based on the form of $\frac{dN(>M)}{dz}$ (the number
of clusters per unit redshift above a mass $M$, c.f. Haiman, Mohr \& Holder 2001), the {\em
only} theoretical requirement is a precise understanding of the scatter in mass around the
detection threshold (e.g. X-ray flux limit).

The rest of this paper is organized as follows.
In \S~\ref{sec:sample}, we compile the local, flux-limited, complete
sample.
In \S~\ref{sec:data}, we describe the Suzaku data, and the spectral analysis
method. We also present spectral fitting result for the 14 clusters.
In \S~\ref{sec:t_l}, we fit the $L-T$ relation in two different
ways.
In \S~\ref{sec:xtf}, we describe our procedures for computing the XTF and
present our results. We also make a comparison with previous results.
%
%In \S~\ref{sec:discussion}, we discuss the limitations of our
%approach, as well as possible future improvements.
%
In \S~\ref{sec:conclude}, we briefly summarize our conclusions and
the implications of this work.
Throughout this paper, we adopt a spatially flat universe dominated by
a cosmological constant and cold dark matter (CDM), with the following
set of cosmological parameters: $\Omega_m=0.27$,
$\Omega_{\Lambda}=0.73$ and $H_0=70~{\rm
  km~s^{-1}~Mpc^{-1}}$. 
% These values are consistent with measurements
%by the {\it WMAP} experiment (Spergel et al. 2003; 2007; we include a
%discussion of the sensitivity of our results to the choice of
%$\sigma_8$ below).

%\maketitle

\section{Local Flux-limited Cluster Sample}
\label {sec:sample}

Our sample is derived from the ROSAT Brightest Cluster Sample (BCS) (Ebeling et al 1998). The BCS
is based on a pure X-ray flux selection criterion using the ROSAT all-sky survey data.
In the northern celestial hemisphere the BCS is (in parallel with the REFLEX survey in the south, Boehringer et al. 2001) the most complete, flux limited sample of low redshift clusters to-date.

In Figure 1 we plot the luminosity versus redshift for BCS
clusters; the flux limit is $4.4\times 10^{-12}$ erg s$^{-1}$ cm$^{-2}$ (0.1-2.4 keV) and all objects have Galactic latitudes $|b|\geq 20^{\circ}$, and redshifts $z\lsim 0.3$. For the
local cluster sample proposed here we have chosen a subset of the BCS with a flux limit of $1 \times 10^{-11}$ erg
s$^{-1}$ cm$^{-2}$ (0.1-2.4 keV) and a redshift limit of $z=0.1$ (shown in Figure 1). Making this flux and redshift cut maximizes the number of objects that have prior X-ray pointed imaging  spectroscopy data while minimizing the number of new datasets required to complete a statistically useful sample. For example, the flux limit required to ensure a 100\% completeness in preexisting spectroscopic data is  $\sim 2.3\times 10^{-11}$ and would result in a sample of only 20 clusters, with only 3 of these with $L_x \leq 10^{44}$ erg s$^{-1}$, compared to 18 below this luminosity in our sample. Our chosen redshift constraint ensures that our
sample is in the local universe, minimizing any redshift dependent variation/scatter
in the L-T relation and XTF which are to be calculated. We note that our flux limit is slightly lower than the
$2\times 10^{-11}$ erg s$^{-1}$ cm$^{-2}$ limit of Ikebe et al (2002, hereafter I02), although their fluxes were recomputed from the ROSAT data and are therefore slightly different (see also Reiprich \& Bohringer 2002). Our final sample contains a total of 61 objects (see below), 14 of which we have observed in  {\sl Suzaku} pointed observations (\S3).

\vspace{3\baselineskip}
\section{Suzaku Data and Analysis}
\label {sec:data}

{\sl Suzaku} is Japan's fifth X-ray astronomy mission, and is equipped with an 
X-ray Imaging Spectrometer (XIS) and hard X-ray detector (HXD). The 14 clusters
identified as having no existing pointed X-ray imaging spectroscopy data were observed using
the Suzaku XIS over an approximately 12 month period from 2006-2007. Table 1 summarizes
these observations.

The Suzaku XIS  consists of four co-aligned X-ray
CCDs (Koyama et al. 2007). Three of these are
front-illuminated (XIS0,2,3), the other is back-illuminated
(XIS1). The CCDs are mounted at the focus of four independent X-ray
telescopes (XRT). Each CCD is $1024\times 1024$
pixels, and covers an image region of $18'\times 18'$ on the sky. The PSF of the XRT has a 
half power diameter of ~$2'$. Sensitivity is in the 
energy range of 0.2-12 keV. 

Exposure times were chosen to ensure that approximately 15,000 XIS counts in {\em
total} (3 FI $+$ 1 BI) were obtained in the cluster core region for
each object. A ratio of flux in the core to the total flux
(integrated to infinity for a $\beta=2/3$ mean emission profile) is a constant 29\%. The
estimated core flux from the BCS was then further reduced by 50\% to allow for vignetting of a diffuse source across the XIS f.o.v.

Not all of the 14 clusters were observed by all 4
cameras, 5 clusters (Z8852, A2665, A2495, A2249 and A566) were
not observed by XIS2. We show XIS0 (raw) images of the 14 clusters in Fig. \ref{fig:img}
together with brightness contours. 
The color and contour are both on a logarithmic scale, and the data are
binned by a factor of 4, then smoothed by gaussian with $\sigma=5$ pixels.

We start our spectral analysis from cleaned event files,
using the standard cleaning process (selection
of the event grade and bad CCD column, disposal of
non-observational intervals, and removal of hot and flickering
pixels). We then accumulate the spectra within a radius of 7.2
arcmin of the nominal cluster center (the peak of surface brightness) using the ``xselect'' Ftool. The spectrum is rebinned by a
factor of 8 so that there are sufficient events in every energy bin for robust fitting.
Blank sky spectra are used as the background, and subtracted from
the cluster spectrum. The net exposure time of blank sky spectra is 96.1 ks. The galactic foreground emission is ignored. This is justified by the fact that our clusters are all hot ($T>3$ keV as we will see below), and the galactic latitudes of Suzaku clusters and blank sky are all over 20 degrees. Thus the amount of galactic emission over the relevant energy range (i.e. where the largest photon count is) is very small in comparison. We  generate the Redistribution Matrix File (RMF) and Ancillary Response File
(ARF) by using the tools  ``xisrmfgen'' and ``xissimarfgen'' included in
the HEASoft package (Ishisaki et al. 2007).  The XIS image is used as the input surface brightness distribution for ``xissimarfgen''. The HEASoft package used is HEAsoft 6.3.0. Note that we use a fixed angular size for accumulating spectra, i.e. higher redshift clusters are sampled with larger physical radius. Recent observations indicate that cluster temperature profiles generally decline with radius outside the core (Vikhlinin et al. 2005,  Leccardi \& Molendi 2008), this implies that in our analyses hotter regions are weighted preferentially for lower redshift clusters. We therefore adopt a phenomenological cluster model that is a reasonable, albeit approximate, fit to currently observed temperature profiles in order to estimate this uncertainty (Reid \& Spergel 2006, Shang et al. 2008 in preparation). We find that the mean emission weighted temperature within 7.2 arcmin of a cluster center and that outside the core ($\sim$ 0.05-0.1 virial radius) differ up to 4\% for z=0.0341 (the lowest redshift in our sample) and z=0.0980 (highest redshift in our sample). The deviation of these temperatures from the mean temperature within $R_{500}$ is also smaller than 4\%. While there may be an effect we therefore consider it to be smaller than other sources of experimental uncertainty - either systematic or random.

%a plot of example spectrum
Ideally the measured intracluster gas temperature would correspond to the virial temperature of cluster.
However, clusters are not always isothermal. Merging, cooling, and phenomena such as AGN outflows  can disrupt isothermality. Approximately half of all systems exhibit evidence for being cooling core clusters (CCC), which have a measurably cooler gas component in their central regions (e.g. Peterson et al. 2003). The cooling core can
contribute a significant fraction of the luminosity of cluster, so
a simple emission-weighted mean temperature may be considerably lower than the true
virial temperature. Excision of the central emission is non-trivial in
the {\sl Suzaku} data due to the PSF. A more appealing solution is to adopt a suitable
spectral model.

In our analyses, we test both single temperature (1T) and two temperature
(2T) models as fits to the net spectrum. A 1T model should provide a better fit to non-cooling core, isothermal clusters (NCCC), while a 2T model should offer an improved fit to a CCC. The 2T model represents 2 independent gas components: a cool component from the core and 
a  hotter component which should be more representative of the virial temperature.

The MEKAL model is taken as our plasma code, as
implemented in XSPEC version 12.3.1x. We simultaneously fit the
spectra accumulated from the different CCD cameras. The energy
range used is restricted to 0.5 keV to 7 keV to mitigate some of the contamination
issues at lower energies and to minimize the background at higher energies relative to the cluster emission. We also tried ignoring the energy band around the Si K-edge (1.825-1.840 keV) due to a potentially anomalous response in that band, however the difference in the resulting temperature is negligible ($<1\%$). The temperatures are tied
together, wheras the normalizations are treated independently for different
cameras (however, the ratios between the normalization of the hot
component and that of the cool component in the 2T model are also tied). We try both
 fixing the metal abundance and setting it as a free parameter. When the abundance is fixed,
the value is set to be 0.3 of solar.
The galactic HI column density is fixed at the value obtained from the LAB survey
(Kalberla et al. 2005). We also enforce
an initial upper limit to the hot component:$T_{hot}$ in our 2T model, since it is assumed to represent the
temperature of the majority of the ICM gas.
%The upper limit is set to be twice of temperature from single temperature fit. 
The 
results of the spectral fits are summarized in Table \ref{tbl:temperature} (metallicity fixed) and Table
\ref{tbl:metallicity} (metallicity as a free parameter). Example spectra are also shown for
 clusters A1800 and Z8276, together with 1T model fit (metallicity fixed) in Fig. \ref{fig:spec}. The fits are good overall, except for the low energy range ($<1$ keV) and some residual Fe features around 1 keV and 6.7 keV for Z8276. We suspect the mismatch below 1 keV is due to uncertainty in contamination in the optical blocking filter of Suzaku XIS (Sato et al. 2007). We estimate this uncertainty by fitting the spectrum using the energy range 1 keV-7 keV, and comparing the result with that from 0.5 keV - 7 keV. For the majority of our clusters, the temperature difference is small ($\sim 2-3\%$), but for a few such as UGC03957 and A2495, the best fit temperature could drop by up to 7\%. The remaining Fe feature in Z8276 appears to be because we fix the metallicity to 0.3 $Z_\odot$, while the true value must be higher than this. This feature disappears when we thaw the metallicity. The resulting temperature is only slightly altered (see Table \ref{tbl:temperature} and Table
\ref{tbl:metallicity}). We also tried replacing the mekal model with the vmekal model, and allowing He and Fe abundance to vary, again the difference in temperature is very small ($<1\%$).

To choose the most likely virial temperature of the cluster from our 1T and 2T models, we perform an F-test that indicates which model is more applicable for the cluster by comparing $\chi^2$ and the number of free parameters..
If including another temperature component (2T model) gives rise to a more than 99\% significant improvement, the  higher temperature ($T_{hot}$) from the
2T model is chosen as the temperature for the cluster, which also implies it is a CCC.
 Otherwise, the 1T model temperature is picked as representing the virial temperature, and
the cluster could be viewed as NCCC. However, there is one system: A2665, where although
the 2T model passes the F-test, we still choose the result from 1T model. In this case $T_{hot}$
from the 2T model for A2665 is more than 3 times the temperature obtained from the 1T model fit, and the normalization of the hot component 
is much smaller than that of cool component. For a typical CCC, it should be larger or at least
of the same order. So we do not believe that $T_{hot}$ is actually representive of the virial temperature for A2665. The model chosen for the virial temperature  is indicated  in bold face in Table \ref{tbl:temperature} and 
Table \ref{tbl:metallicity}. The temperatures shown in Table \ref{tbl:temperature} are used in our analyses below. 

Comparing the results in Table \ref{tbl:temperature} and Table \ref{tbl:metallicity}, we see that the temperature generally
varies only a little after setting the metallicity as a free parameter instead of
fixing it at 0.3 solar. The one exception is UGC03957. This system is better fit as a NCCC after we thaw the
metallicity parameter, it also has the highest metal abundance in our {\sl Suzaku} survey (0.59 $Z_{\odot}$). 

To check whether the {\sl Suzaku} temperatures obtained using our procedure are consistent with published value from other missions, we also performed our
analysis on 3 clusters (A1060, A2218 and A1795) with well established temperatures in the literature from multiple missions (e.g. ASCA, Chandra, XMM). Our results appear to match
previous results quite well in both NCCC and CCC cases. A detailed comparison is listed in Table \ref{tbl:observation}. Below are a few notes for each cluster.

A1060: With z=0.00114, it is one of the nearest Abell clusters. Various observations show that it has uniform temperature and abundance distribution, and is therefore thought to be in a dynamically relaxed state (Tamura et al. 1996, Furusho et al. 2001, Hayakawa et al. 2004, Hayakawa et al. 2006). Although recent observations indicate that the temperature profile has a steep drop in the outer regions (Sato et al. 2007), the variation within 7 arcmin is small. Obviously, a simple 1T model is sufficient and perfectly suitable for such a case. All temperature measurements give quite close results: $3.26\pm 0.06 $keV from ASCA(Furusho et al. 2001), $3.36^{+0.05}_{-0.06}$ keV from Chandra (Hayakawa et al. 2004), $3.27\pm 0.02$ keV from XMM-Newton. Our result is also in perfect agreement with them.

A2218: Residing at z=0.175, this cluster is further away than any object in our sample. Various observations suggest it is at a later merger stage with a centrally peaked and azimuthally asymmetric temperature map (Govoni et al. 2004; Machacek et al. 2002; Markevitch et al. 2000). A 1T model is preferred for this cluster since there is no cooling flow in the center, and our result is in close agreement with previous measurements.

A1795: Is a strong cooling flow cluster at z=0.0634. Different measurements and different treatments of the cooling flow give very different results, varying over a wide range of 5.3-7.8 keV. The low end of the measured temperature range (5.3keV) is from 1T models that do not take into account of cooling flow. But even within the analyses that explicitly considered the cooling flow (fitting with a 2T model, cooling flow model, or excising the cooling flow region), the range of results still appears to be large. Fortunately, recent results seem to stablilize around 6 keV.  Among them, Allen et al. (2001) extracted the temperature from ASCA data using 5 different models. Their result is summarized as follows: $5.40^{+0.08}_{-0.09}$keV from a 1T model (absorbing column density fixed), $5.33^{+0.10}_{-0.11}$keV from a 1T model (absorbing column density as a free parameter), $5.87^{+0.25}_{-0.19}$keV from a constant cooling flow model, $6.21^{+0.25}_{-0.22}$keV from a 2T model, an
 d $5.61^{+0.17}_{-0.15}$keV from a isothermal cooling flow model. An incomplete list of other measurements includes: $6.1$keV from Chandra (Vikhlinin et al. 2005, 1T model with central 70 kpc excluded), $5.95\pm0.1$keV from XMM-Newton (Arnaud et al. 2001, 1T fit with cooling flow region excluded). Our result ($5.73^{+0.23}_{-0.20}$keV is in good agreement with these recent measurements.

The ROSAT fluxes used to define our sample are computed from raw counts
assuming a spectral model and gas temperature derived in an iterative fashion from
an assumed $L-T$ relation (Ebeling et al. 1998). We have therefore checked whether our
spectroscopically derived temperatures alter the cluster flux sufficiently in any case
to cause systems to fall below our flux limit. We find that the fluxes vary only some 1-2 \% even if the temperature
changes by a factor of 2. Since this is small compared to the ROSAT count rate uncertainty
(around 15\%) it is unlikely to bias the sample either by excluding or including systems at
the flux limit. The ROSAT flux is therefore shown in Table \ref{tbl:sample}, but luminosities have been
computed  using our updated temperatures where appropriate.

57 clusters in Table \ref{tbl:sample} have measured/published temperatures,
which is 93\% of our sample. 34 clusters were measured by
ASCA, 14 were measured by Suzaku, 6 by Einstein or EXOSAT, 2 by
Chandra, 1 by XMM-Newton, 4 used estimated temperature from our
$L-T$ relation which is to be calulated. When a cluster has temperature
measurements by several instruments, our first choice is to adopt the
ASCA measurement in order to minimize any instrumental systematics. 
More than half of the clusters in the sample have ASCA observations, 
in addition the ASCA SIS is very similar to the {\sl Suzaku} XIS. Our
second choice is to use Chandra or XMM-Newton for their accuracy,
lastly we choose others. If the spectrum of a cluster was fitted
with different models in the literature, we choose 2T model fits over cooling flow models
to ensure consistency with our analysis.

Our sample differs from that of the HIFLUGCS (Reiprich\&B\"ohringer 2002) used by I02 in the
follwing aspects: HIFLUGCS covers both northern and southern skies, ours
only covers the northern sky; HIFLUGCS uses a flux limit of $2.0\times 10^{-11}\mbox{erg s}^{-1}\mbox{ cm}^{-2}$,
while ours is a factor of two lower; the two samples are constructed based on ROSAT
flux estimated in slightly different ways.

\section{The L-T Relation}
\label{sec:t_l}

The X-ray luminosity-temperature ($L-T$) relation for galaxy clusters has long been used
as a probe of their physics and evolution. In this section, we fit our data above with the following relation,
\begin{eqnarray}
\mbox{log}L_{0.1-2.4\mbox{keV}}(h^{-2}\mbox{ ergs} \mbox{
s}^{-1})=\alpha+\beta\mbox{log}T(\mbox{keV}) \label{eq:t_l}
\end{eqnarray}
The $L-T$ relationship has an intrinsic scatter in addition to any measurement errors. A common way of dealing with this
kind of problem in curve fitting is to employ the BCES($\mathrm{X}_{2}|\mathrm{X}_{1}$)
estimator(Akritas \& Bershady 1996). However, the slope derived from the BCES estimator depends
on the measurement error. In fact, the data with the largest errors/scatter are effectively given more weight in the
fit than the data with smaller errors/scatter, contrary to naive expectations. In an attempt to mitigate this inherent bias we also performed a modified $\chi^{2}$ fit which takes intrinsic scatter into account explicitly. Our $\chi^{2}$ is defined as the following:
\begin{eqnarray}
\chi^{2}=\sum_{i} \frac{(\mbox{log}L_{i}-\alpha-\beta\mbox{log}T_{i})^{2}}{\sigma^{2}_{\mbox{log}L_{i}}+\beta^{2}\sigma^{2}_{\mbox{log}T_{i}}+\sigma^{2}_{\mbox{intr}}}
\label{eq:chi}
\end{eqnarray}
where $\sigma^{2}_{\mbox{intr}}$ is the intrinsic scatter, and
$\sigma^{2}_{\mbox{log}L_{i}}$ and $\sigma^{2}_{\mbox{log}T_{i}}$
are the measurement variances in $L$ and $T$. $\sigma^{2}_{\mbox{intr}}$ is
chosen so that the minimum reduced $\chi^{2}$ is 1. The 
results of both a standard BCES fit and that of our modified $\chi^{2}$ are listed in Table \ref{tbl:fit}. 

As we can see, the two methods are roughly consistent for
our data, but the $\chi^{2}$ fit has the advantage that small error
measurements are given more weight in the fitting. This approach yields
slightly lower power law indices for the $L-T$ relation.
 
In the analyses described
below, we use the result from the $\chi^{2}$ fitting. The
fits together with data are plotted in Fig. \ref{fig:data}.

 In the same way we also fit the relation between bolometric luminosity and temperature. These results are also presented in Table \ref{tbl:fit}, the measured slope steepens to $\sim 3$. The conversion between band and bolometric luminosity is computed using the Mekal plasma code and measured temperature. Compared with high redshift measurements (e.g. Ota et al. 2006), the slopes match within the uncertainty, while our normalization is smaller, which is in general agreement with the positive redshift evolution of luminosity that has been suggested by previous works (Vikhlinin et al. 2002, Lumb et al. 2004, Kotov \& Vikhlinin 2005).
\section{X-ray Temperature Function}
\label{sec:xtf}
\subsection{Derivation of XTF}
\label{subsec:xtf}

As a first step towards using our cluster sample to provide a complementary zero-redshift
calibration point for cosmological studies (alongside, for example, Reiprich \& B\"ohringer 2002, Ikebe et al. 2002) we have made a preliminary assessment of the cluster X-ray temperature function (XTF). 

The number density of clusters hotter than a
 temperature $T$ is given by,
\begin{eqnarray}
n(T)=\sum_{T<T_{i}}\frac{1}{V_{\mbox{max}}(T_{i})} \label{eq:xtf}
\end{eqnarray}
where $V_{\mbox{max}}(T)$ is the maximum comoving volume where a
cluster of temperature $T$ would be detected. Following many previous
works, we assume that $\mbox{log}L$ follows a normal distribution
with mean value given by the $L-T$ relation above, and with
variance $\sigma_{\mbox{intr}}$.
\begin{eqnarray}
V_{\mbox{max}}(T)=
 \int_{-\infty}^{+\infty}
\frac{v_{\mbox{max}}(T,L)}{\sqrt{2\pi\sigma^{2}_{\mbox{intr}}}}\mbox{exp}\left(-\frac{(\alpha+\beta\mbox{log}T-\mbox{log}L)^{2}}{2\sigma^{2}_{\mbox{intr}}}\right)d\mbox{log}L
\label{eq:vmax}
\end{eqnarray}
where the $v_{\mbox{max}}(T,L)$ is the maximum search volume for a
cluster with temperature $T$ and luminosity $L$.
\begin{eqnarray}
v_{\mbox{max}}(T,L)=\frac{\Omega}{3}D_{\mbox{com}}^{3}(z_{\mbox{max}})
\label{eq:vsm}
\end{eqnarray}
where the $\Omega$ is the solid angle sky coverage of the survey (equal to 4.14 for the ROSAT BCS), and
$z_{\mbox{max}}=Min(z_{f},z_{s})$. $z_{s}$ is the redshift limit
of our sample, which is 0.1. $z_{f}$ is constrained by the flux
limit, and is computed by solving the following equation,
\begin{eqnarray}
4\pi(1+z)^{2}D^{2}_{\mbox{com}}(z_{f})f_{\mbox{lim}}=L_{\mbox{obs}}(L,T,z_{f})
\label{eq:zf}
\end{eqnarray}
where $L_{\mbox{obs}}(L,T)$ is the observed luminosity after applying a
K-correction, and $f_{\mbox{lim}}$ is the flux limit of our sample,
which is $1.0\times10^{-11}\mbox{erg}\mbox{ cm}^{-2}\mbox{
s}^{-1}$. Since we assume a flat universe model, the comoving
distance takes a simple form,
\begin{eqnarray}
D_{\mbox{com}}(z)=\frac{1}{\mathrm{H}_{0}}\int_{0}^{z}\frac{1}{\sqrt{\Omega_{m}(1+z^{\prime})^{3}+\Omega_{\Lambda}}}d
z^{\prime} \label{eq:dcom}
\end{eqnarray}

The calculated $V_{\mbox{max}}(T_{i})$ is listed in Table
\ref{tbl:sample}. We plot the cumulative XTF  in Fig. \ref{fig:xtf} together with
the results from previous works. We find general agreement with earlier measurements at temperatures 
above $\sim 5$ keV. At lower temperatures we see a slightly lower normalization than either the I02 or Henry (2000) measurements. 

A particular feature in our cluster sample, that is also evident to some extent in the I02 and Henry (2000) samples, is a small gap or deficit of clusters in the temperature range of $\sim 4.5-5.5$ keV. 
This is also apparent in Figure 4 and in Figure 6 where we plot cluster temperature versus redshift for our sample. In particular, the {\sl Suzaku} measurements appear
grouped towards 4-4.5 keV across a range in redshift. Although this may suggest a systematic issue in our temperature
estimates we have also plotted in Figure 6 the results of our identical analysis of  {\sl Suzaku} data on the archived objects A1060, A2218, and A1795 (Table 4), which agree very well with pre-existing measurements from other missions, and show no obvious biases. This gap is also apparent in the pre-existing temperature measurements that make up the rest of our sample. Although the I02 and Henry 2000 cumulative XTF's also show a similar feature it is notable that the Markevitch (1998) XTF does not. All samples have very similar redshift limits (i.e. $z<0.2$ with very few objects beyond $z=0.1$). Markevitch (1998) derives ASCA temperatures by explicitly removing emission from any cool cores (previously ascribed to large cooling flows), whereas I02  and our present work employ a similar 2T/1T spectral approach and Henry (2000) used many collated temperatures. It is therefore possible that systematic effects in temperature measurements are responsible for this feature. We do how
 ever believe that our 1T/2T modeling approach should avoid any bias towards {\em underestimating} temperatures, since we do detect and reduce the contribution of any cool cores.

It is of course possible that this feature is simply a statistical fluke, due to the finite sample size and $L-T$ scatter. Indeed, there is a clear steepening in the cumulative XTF slope (Figure 5) at temperatures immediately below 4.5 keV compared that above 5.5 keV, indicating a slight ``pile-up" of objects cooler than 4.5 keV. If we naively average over the temperature range 2-8 keV then there are approximately 4.5 clusters per 0.5 keV bin. Finding zero clusters in the 4.5-5 keV range (see Table 5) is therefore (assuming Poisson statistics) significant at about a $2-\sigma$ level.  Finding 10 clusters in the 4-4.5 keV range (see Table 5) is a positive deviation at about a $2-3\sigma$ level. Taken together this suggests that while a statistical fluctuation cannot be ruled out, it would certainly be well outside a 2-$\sigma$ event.

It is also possible that this is a manifestation of cosmic variance due to large-scale structures within the finite volume of our sample (which is distinct from a pure statistical fluke). This is something that has plagued very early efforts to make cosmological methods with galaxy clusters (e.g. Edge et al. 1990). The apparent dearth of objects at $z\sim 0.05$ (Figures 1 and 6) is likely a consequence of this, although this does not obviously map to the deficit of 4.5-5.5 keV clusters.

\section{Summary and Conclusions}
\label{sec:conclude}
 
 We have conducted an X-ray galaxy cluster survey of 14 systems using {\sl Suzaku}. These clusters are part of a  statistically complete sample of 61 clusters in the local ($z<0.1$) Universe with $f_x>1.0 \times 10^{-11}$ erg s$^{-1}$ cm$^{-2}$ (0.1-2.4 keV), in the northern celestial hemisphere. None of these 14 clusters have been observed previously in pointed X-ray observations, despite being luminous systems ($L_x>1 \times 10^{43}$ erg s$^{-1}$ (0.1-2.4 keV)). The {\sl Suzaku} data is shown to be of high quality and enables robust imaging and spectral analyses to be made. A comparison of spectral analyses of {\sl Suzaku} data on 3 well known clusters (not in our sample) with those from earlier missions indicates no obvious biases either in the {\sl Suzaku} calibration or the analysis approach that we use.
  
We present the first results of a spectral analysis of these 14 clusters with the aim of deriving gas temperatures that reflect the true virial temperature. To do this we employed a spectral modeling scheme that determines whether a one or two plasma temperature model is appropriate. Two temperature models allow us to separate out the contribution of cooling cores in 7 of the clusters. Combining the new {\sl Suzaku} data with preexisting temperature data for the flux limited sample we obtain a dataset of 57 objects with spectroscopically confirmed gas temperatures and 4 objects for which pointed X-ray data is either not yet public or has not yet been analyzed. For these 4 systems we simply use estimated temperatures based on the $L-T$ relation, and note that their inclusion or exclusion has a negligible impact on our results.

 We have determined both the $L-T$ relation and the cumulative XTF for the full sample. In fitting the $L-T$ relationship with a power law we argue that the often used BCES method can produce a power law index that is dependent on the actual measurement errors in $L$ and $T$, owing to the weighting scheme employed. We have therefore also employed a modified $\chi^2$ fit that takes intrinsic scatter into account in an attempt to mitigate any bias in the fit. We find that  this modified $\chi^2$ and the BCES  method are in quite close agreement, with power-law indices differing by $\sim 4$\% and normalizations by $\sim 0.1$ \%. The modified $\chi^2$ yields $L_{0.1-2.4\rm{keV}}= 10^{42.31} T^{2.45}$. 
 
 The measured XTF is in close agreement with that of previous studies. Obviously there is significant overlap in the actual clusters contained in all such works. However, our lower flux limit (by a factor of 2 from, for example, I02) and the fact that 22\% of our sample has never been studied in the X-ray before, helps reduce this problem.  There is some evidence that our XTF is slightly lower at temperatures less than $\sim 4.5$ keV than the earlier measurements. This appears to be partially a consequence of a deficit of clusters in the range of 4.5-5.5 keV. We have considered several possible causes for this. These include; a systematic bias in our spectral analyses, a statistical fluke, or a consequence of large-scale structure and cosmic variance in this very local sample. Since a similar deficit is also apparent in two earlier XTF measurements (I02, Henry (2000)) it may be that a statistical or cosmic variance effect is the most likely. 
 
If local galaxy cluster samples are to be used to construct robust calibrations for future high-precision cosmological tests then it will be crucial to ensure that we understand the finite volume effects (e.g. cosmic variance) that may produce systematic biases. One solution will be to push measurements to slightly lower mass scales where clusters are more numerous. We have demonstrated that {\sl Suzaku} is an excellent instrument for doing just that, by exploring the population of luminous, nearby, clusters that has nonetheless not yet been fully studied.

\vspace{-0.5\baselineskip}

\acknowledgments We thank Zolt\'an Haiman and an anonymous referee for helpful comments. This work was supported by NASA {\sl Suzaku} grant NNX06AI40G, and we thank the entire {\sl Suzaku} team and the staff of the {\sl Suzaku} Guest Observer Facility.

%\vspace{-2\baselineskip}

\clearpage

\begin{figure}[tbg]
\epsscale{0.1}
\begin{tabular}{c}

\rotatebox{-0}{\resizebox{150mm}{!}{\includegraphics{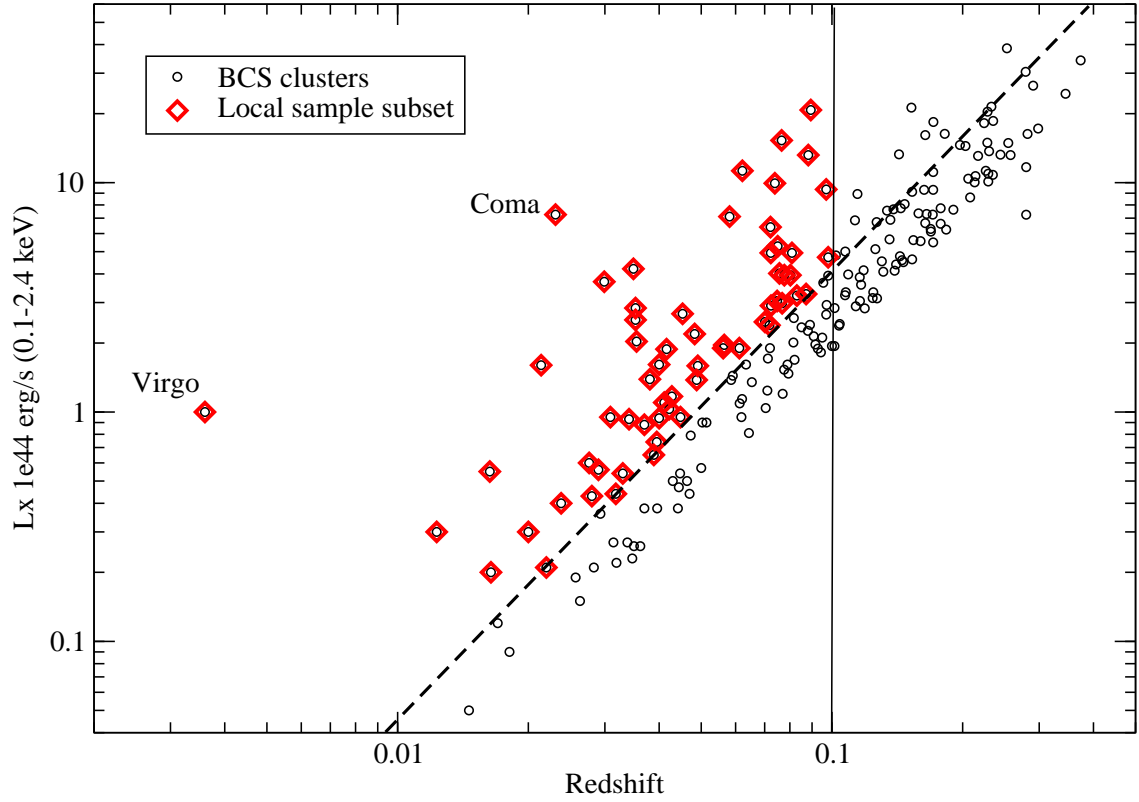}}}
\end{tabular}
\caption{The X-ray luminosity of all BCS clusters is plotted versus redshift. The flux and redshift limits of our sample of 61 objects are shown by the dashed and solid lines respectively. Notice the apparent deficit of luminous objects at $z\sim 0.05$. 
\label{fig:bcs}}
\end{figure}

\begin{figure*}[tbg]
\epsscale{0.1}
\begin{tabular}{cc}
%UGC03957&A2572b&Z8852&IIZw108\\
\rotatebox{-0}{\resizebox{90mm}{!}{\includegraphics{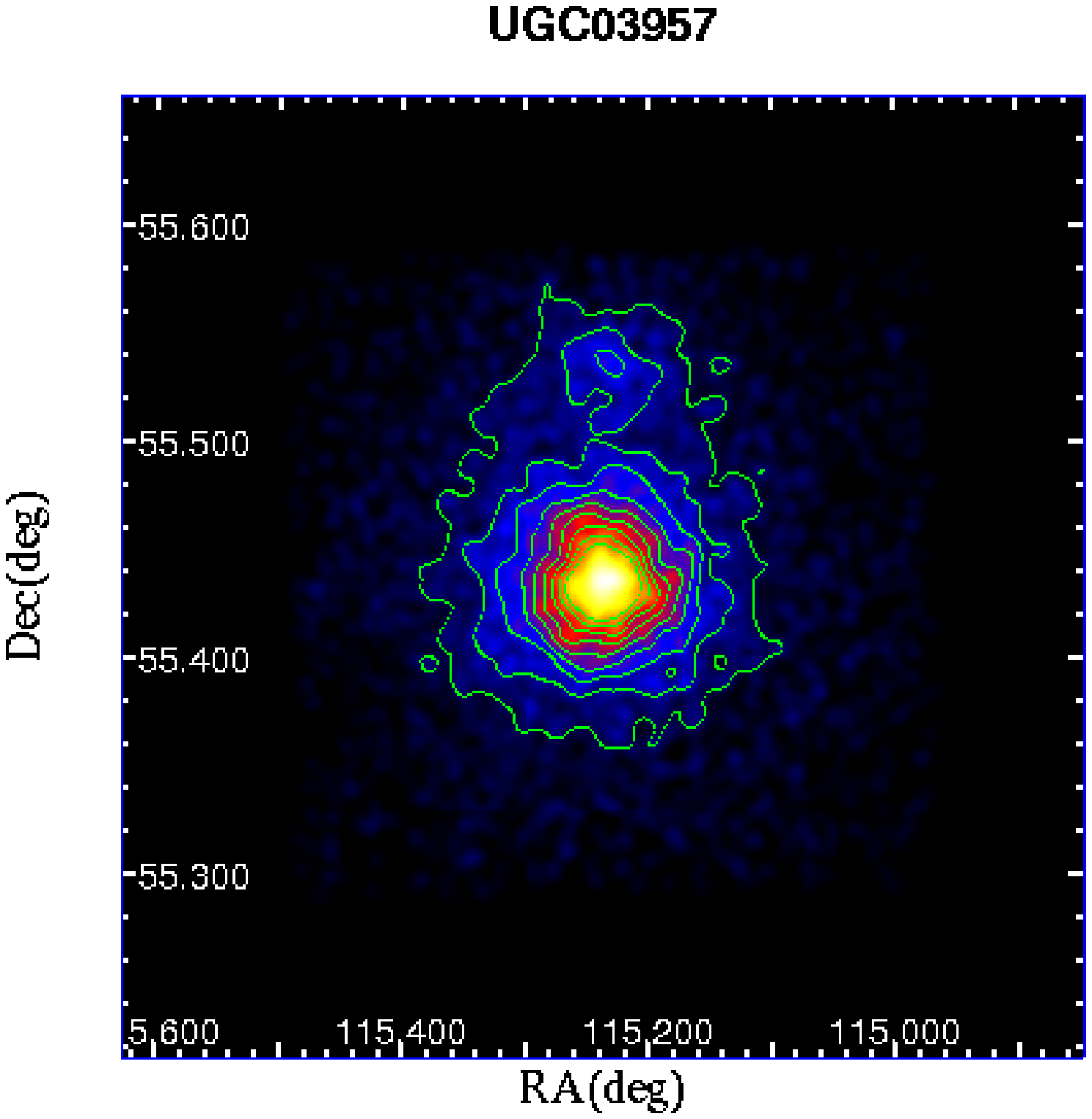}}}&
    \rotatebox{-0}{\resizebox{90mm}{!}{\includegraphics{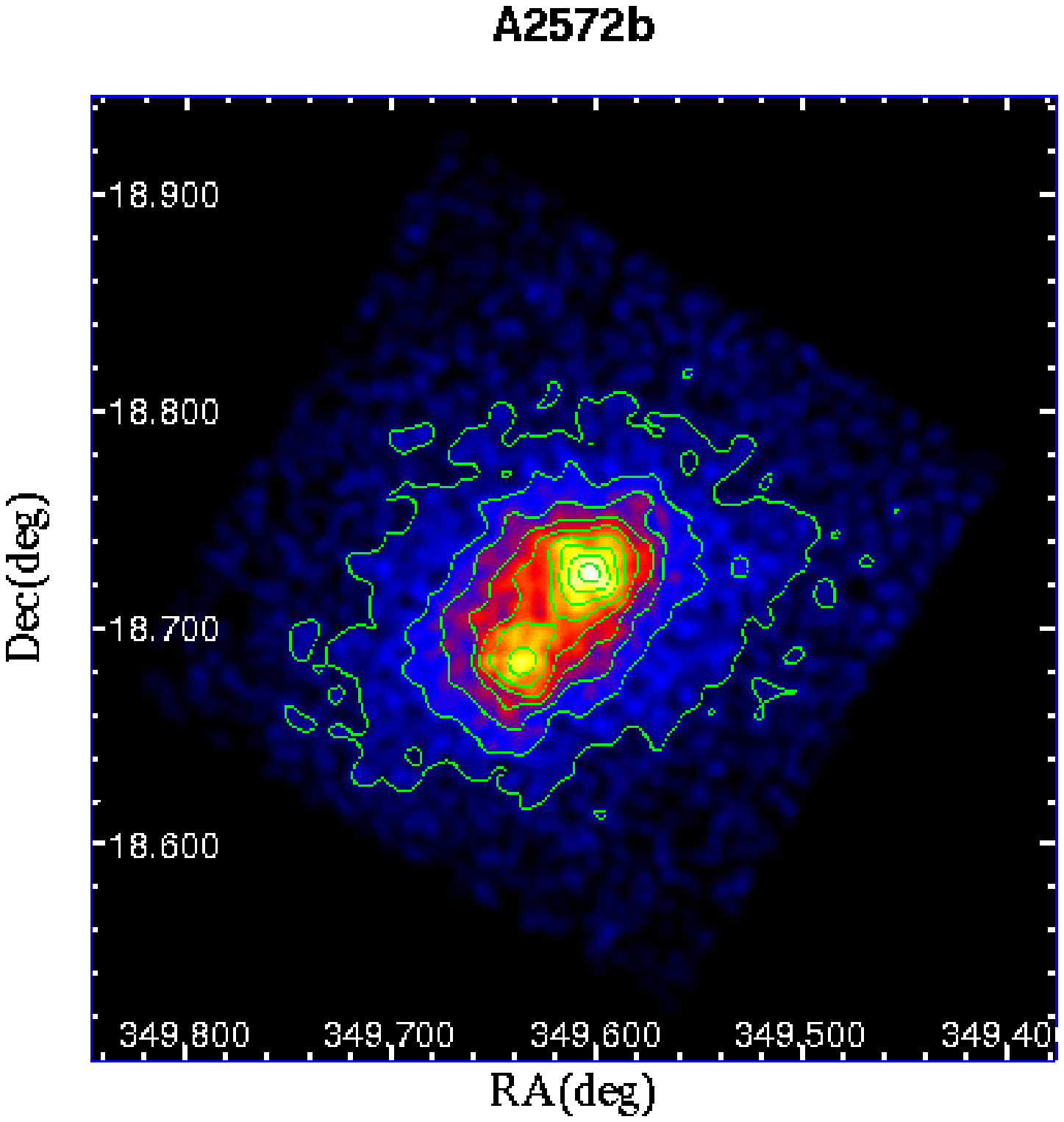}}}\\
\rotatebox{-0}{\resizebox{90mm}{!}{\includegraphics{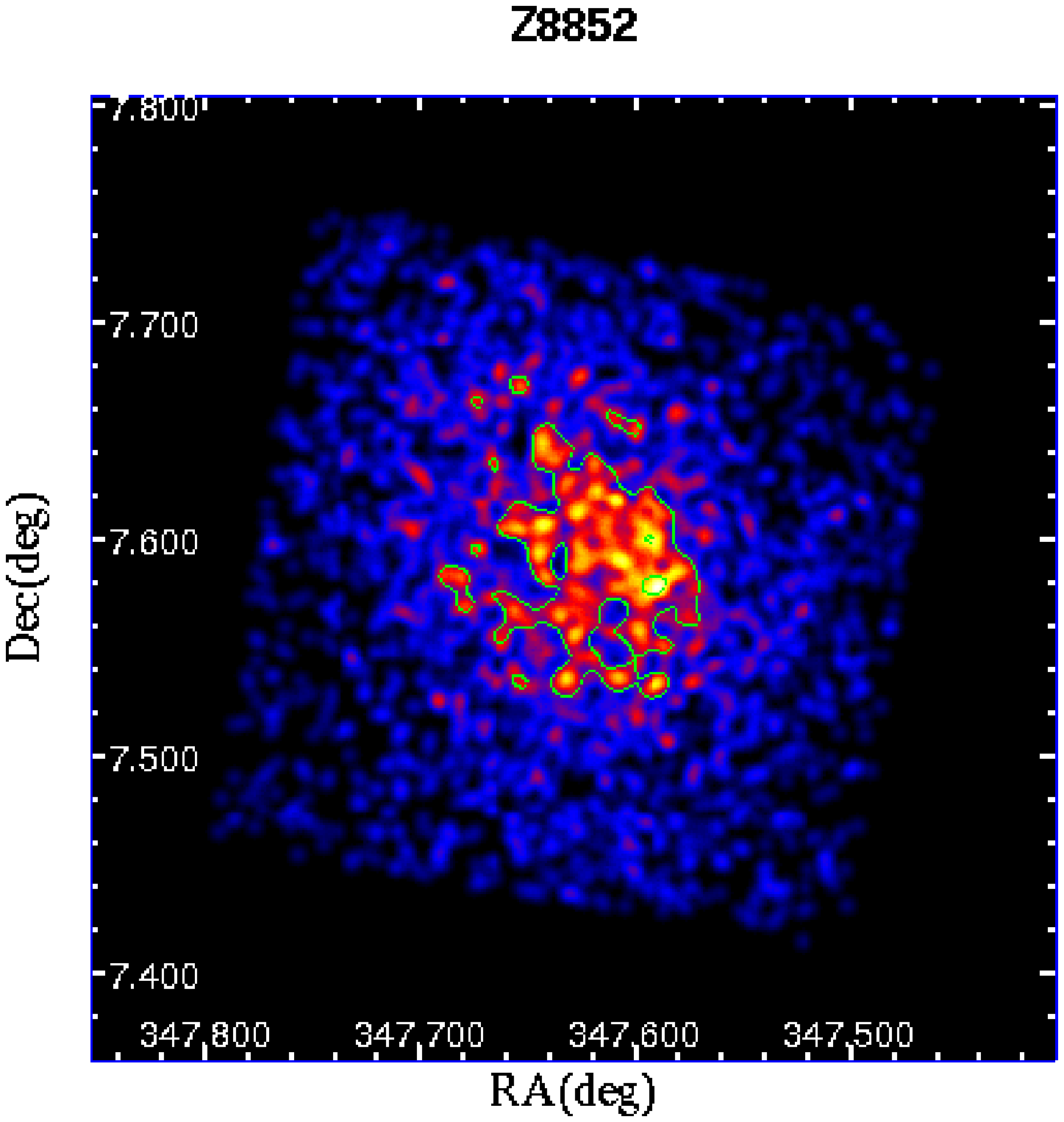}}}&
\rotatebox{-0}{\resizebox{90mm}{!}{\includegraphics{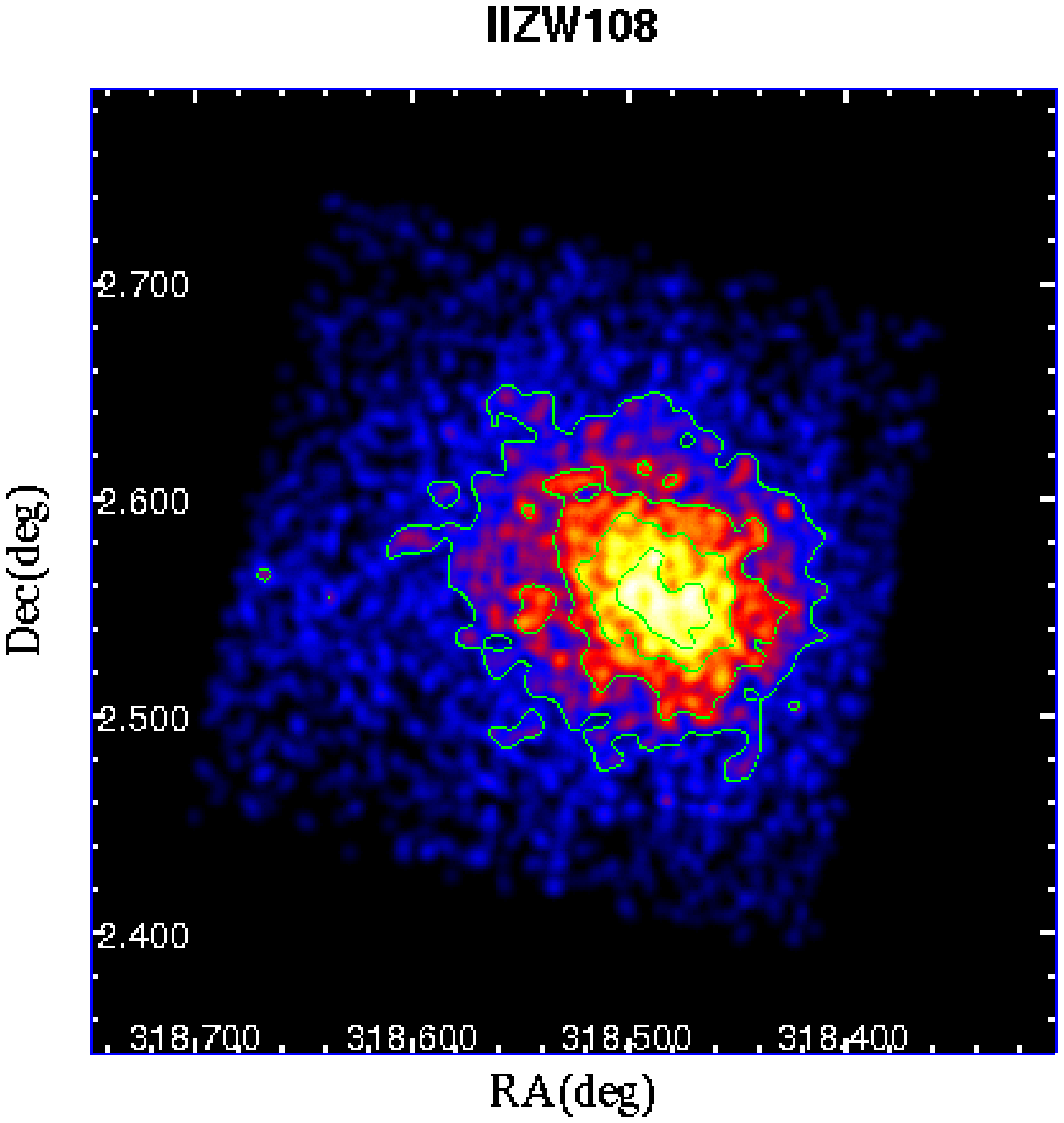}}}\\
%A2665&A1831&A1800&Z8276\\
\end{tabular}
\caption{
Raw  images of clusters (background NOT subtracted) from the XIS0 camera in the 0.5-2 keV energy range
together with brightness contours.
The colour and contours are both on a logarithmic scale. The data were binned by 4, and
smoothed with a $\sigma=5$ pixel Gaussian.
\label{fig:img}}
\end{figure*}

\begin{figure*}[tbg]
\epsscale{0.1}
\begin{tabular}{cc}
%UGC03957&A2572b&Z8852&IIZw108\\
\rotatebox{-0}{\resizebox{90mm}{!}{\includegraphics{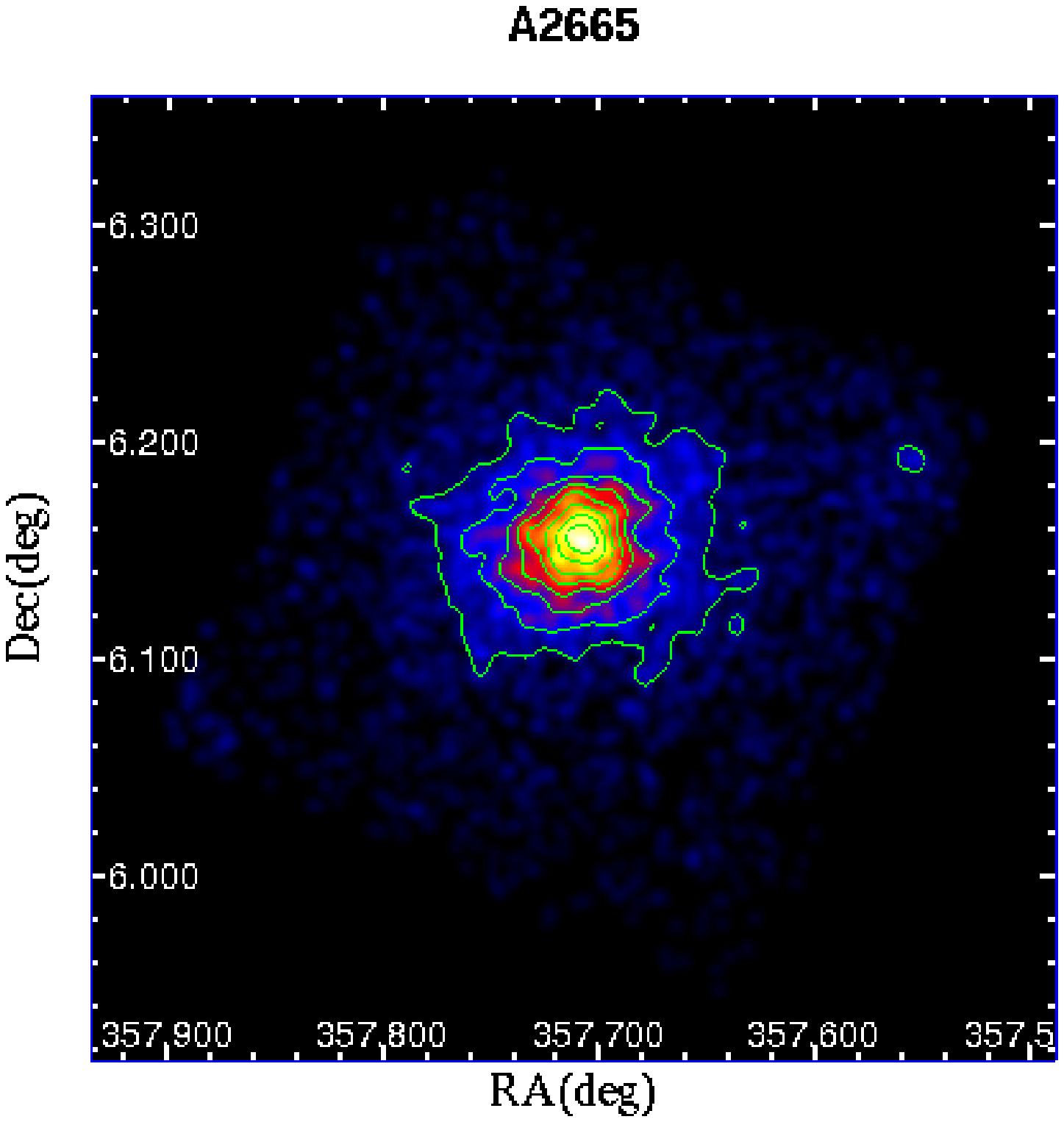}}}&
    \rotatebox{-0}{\resizebox{90mm}{!}{\includegraphics{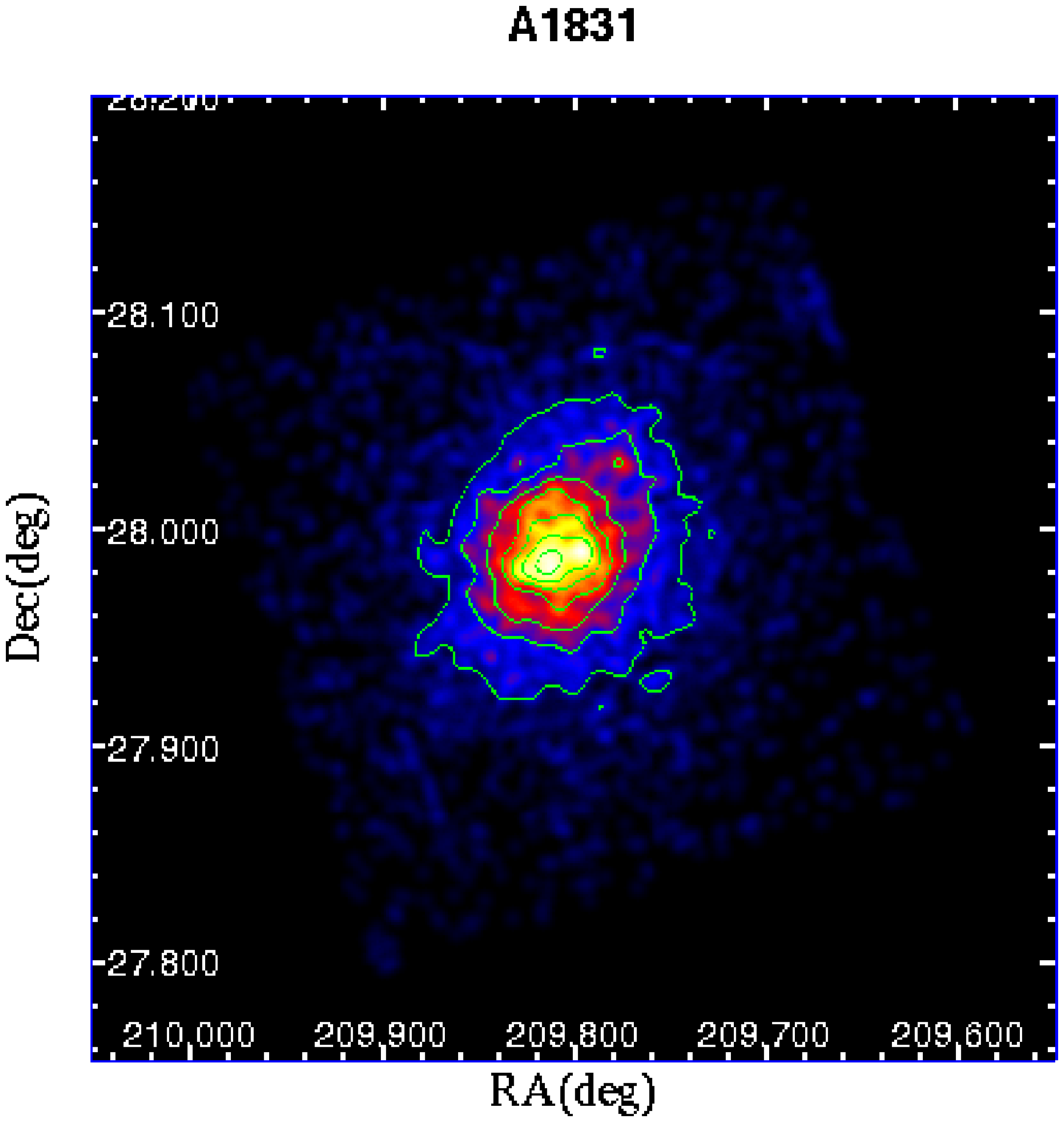}}}\\
\rotatebox{-0}{\resizebox{90mm}{!}{\includegraphics{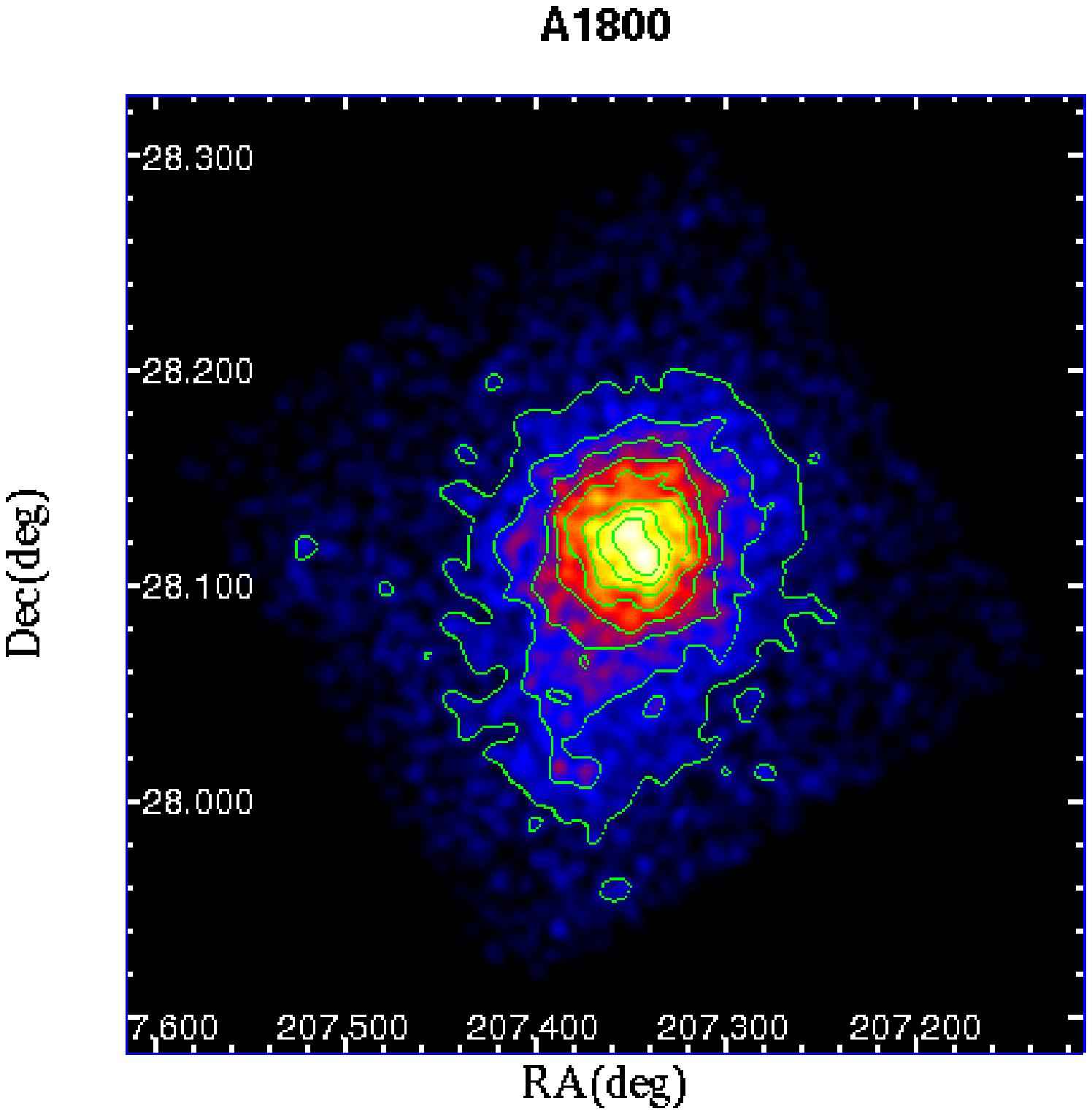}}}&
    \rotatebox{-0}{\resizebox{90mm}{!}{\includegraphics{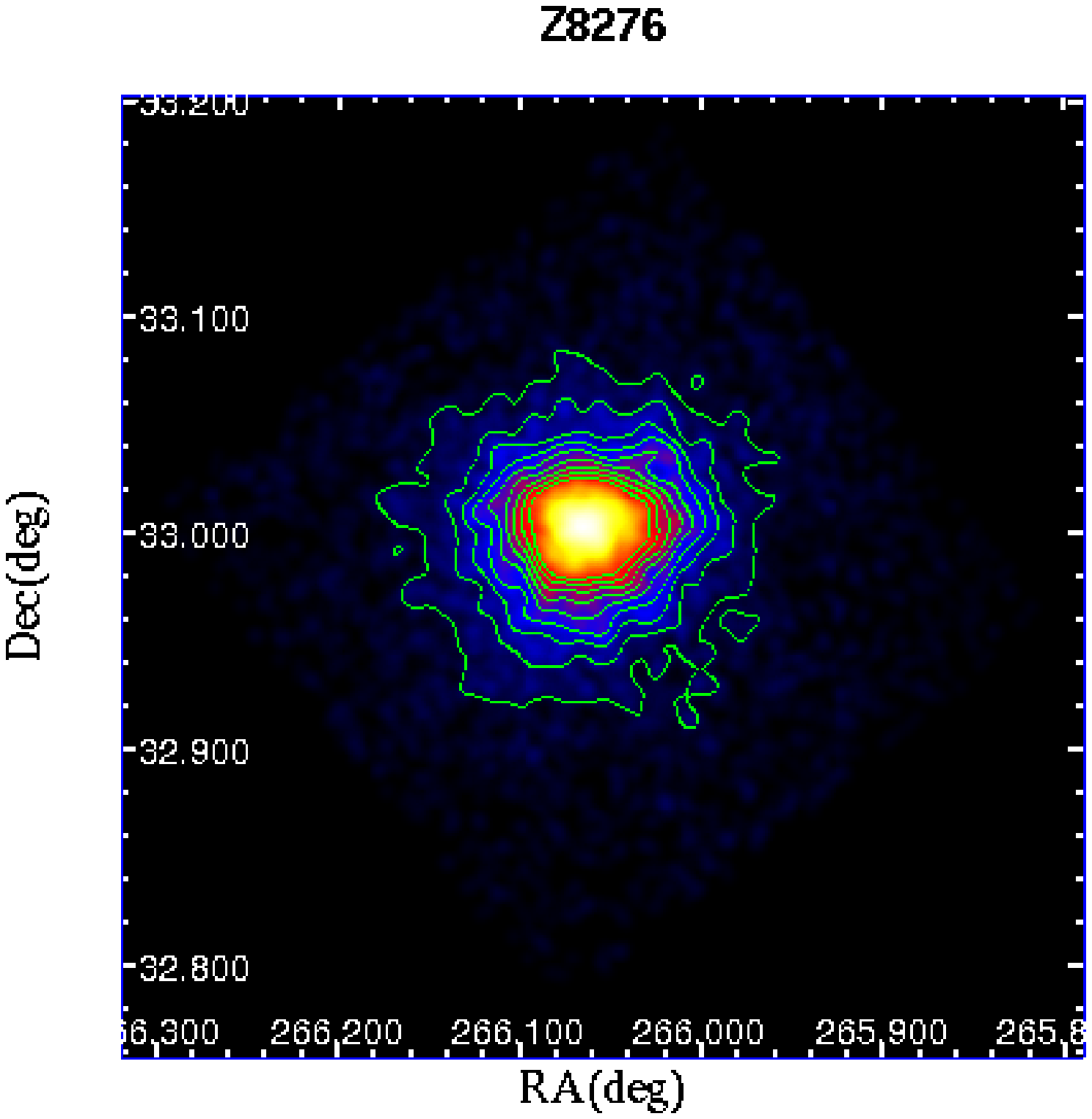}}}\\
\rotatebox{-0}{\resizebox{90mm}{!}{\includegraphics{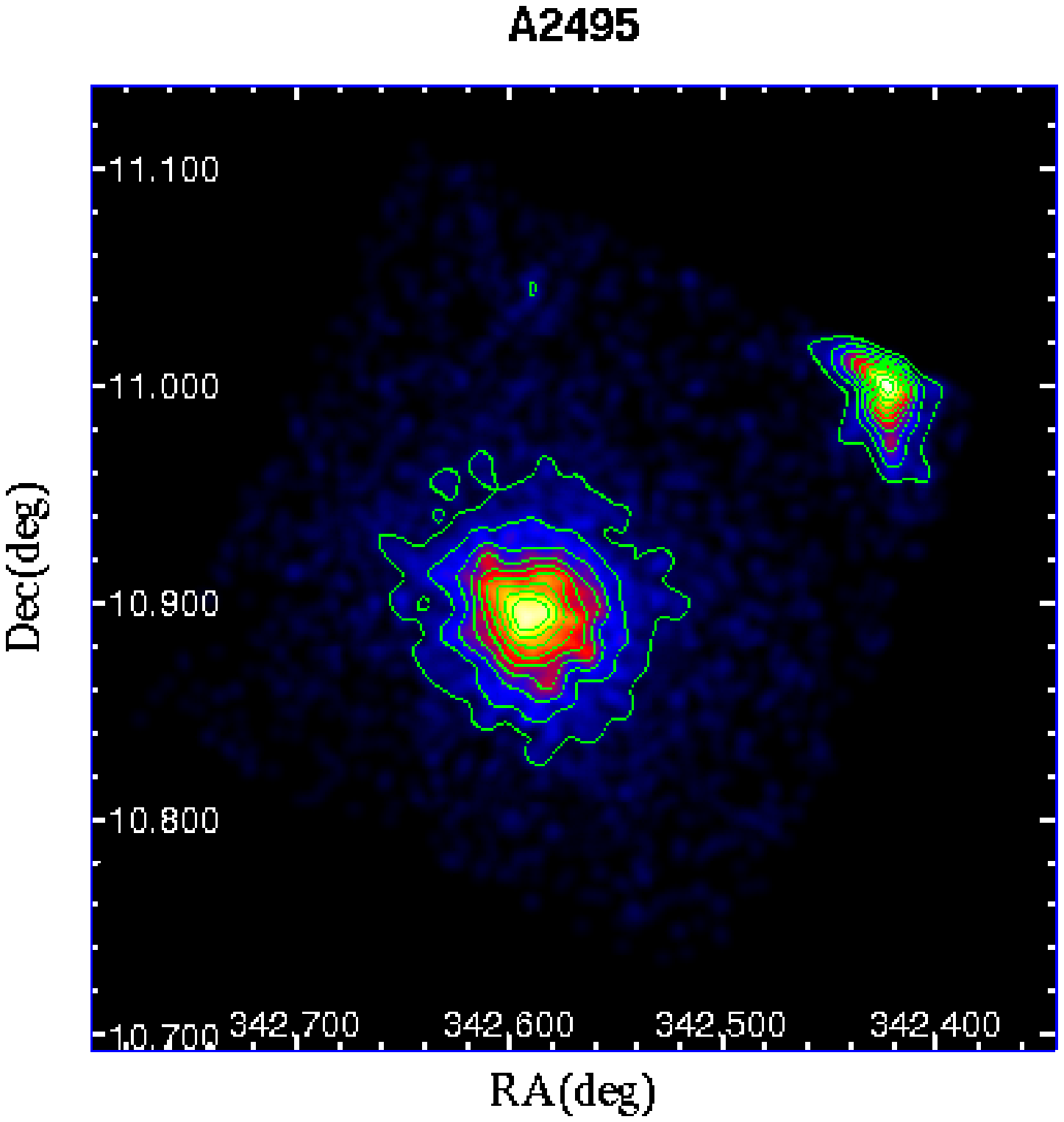}}}&
\rotatebox{-0}{\resizebox{90mm}{!}{\includegraphics{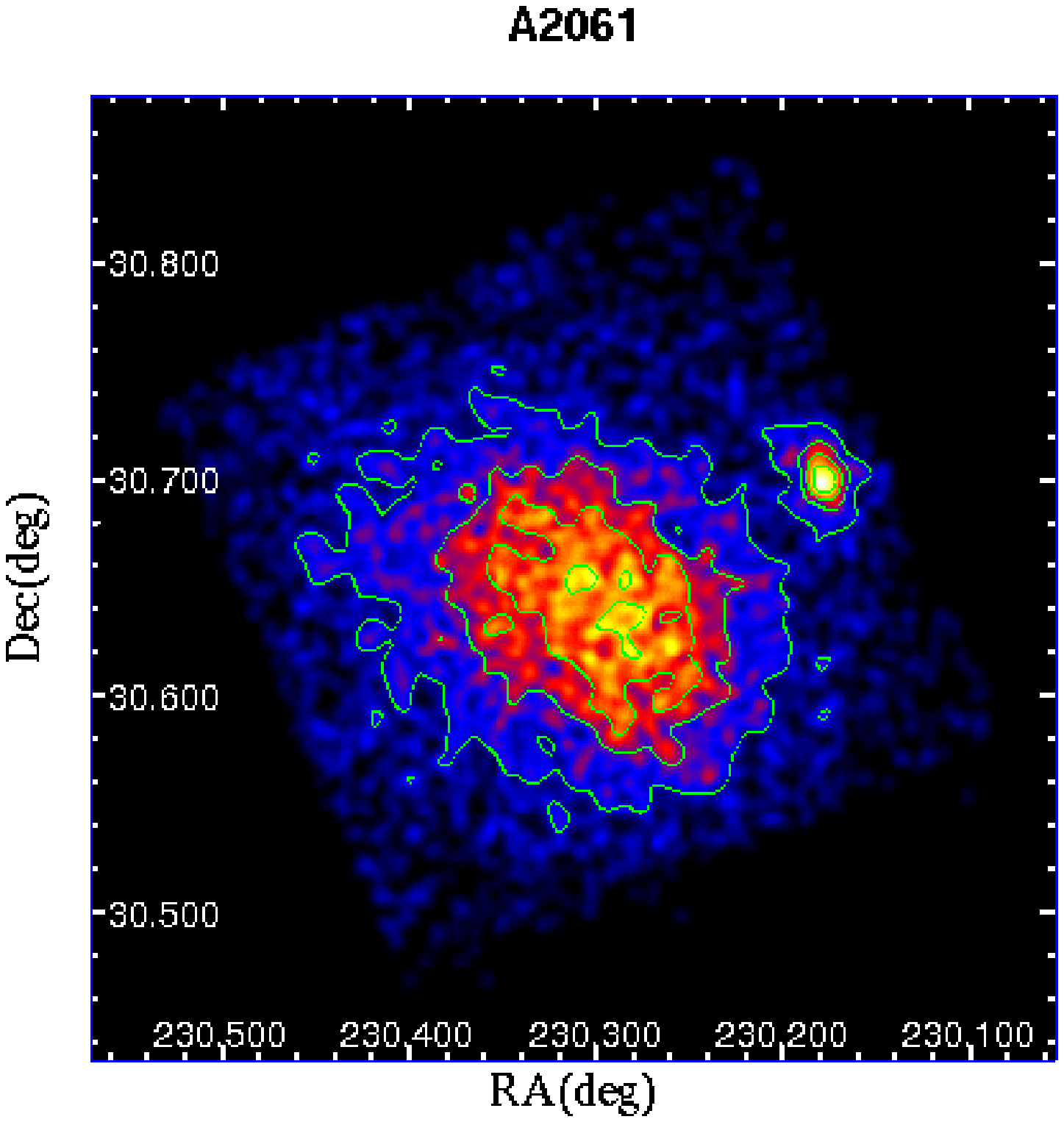}}}\\
%A2665&A1831&A1800&Z8276\\

\end{tabular}
\parbox{0.5\textwidth}{
\begin{minipage}{18cm}
\small\parindent=3.5mm
{\sc Fig.} \ref{fig:img}.--- Continued
\end{minipage}
}
%\caption{Contined.}
%\label{fig:img}
\end{figure*}

\begin{figure*}[tbg]
\epsscale{0.1}
\begin{tabular}{cc}

\rotatebox{-0}{\resizebox{90mm}{!}{\includegraphics{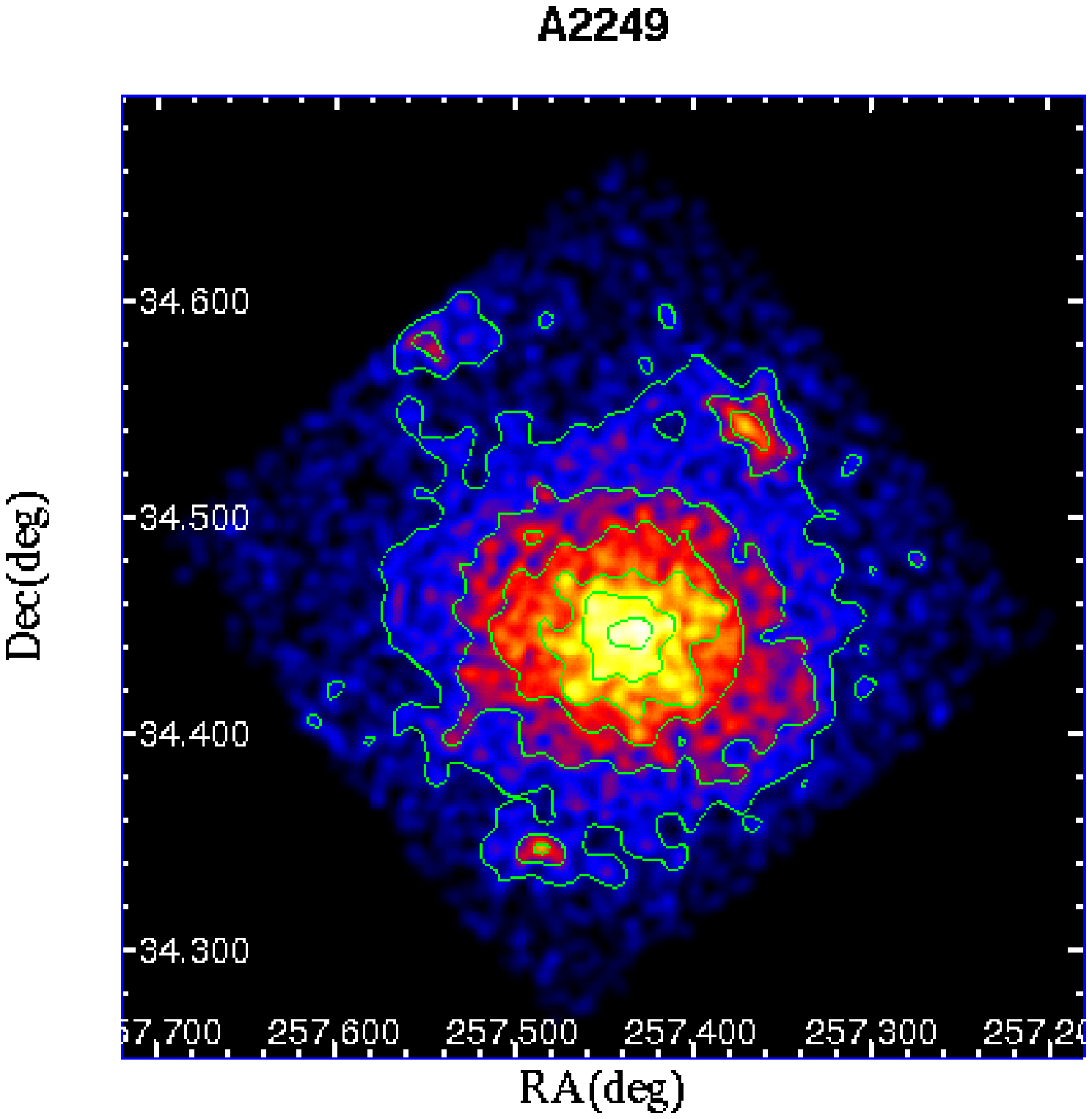}}}&
    \rotatebox{-0}{\resizebox{90mm}{!}{\includegraphics{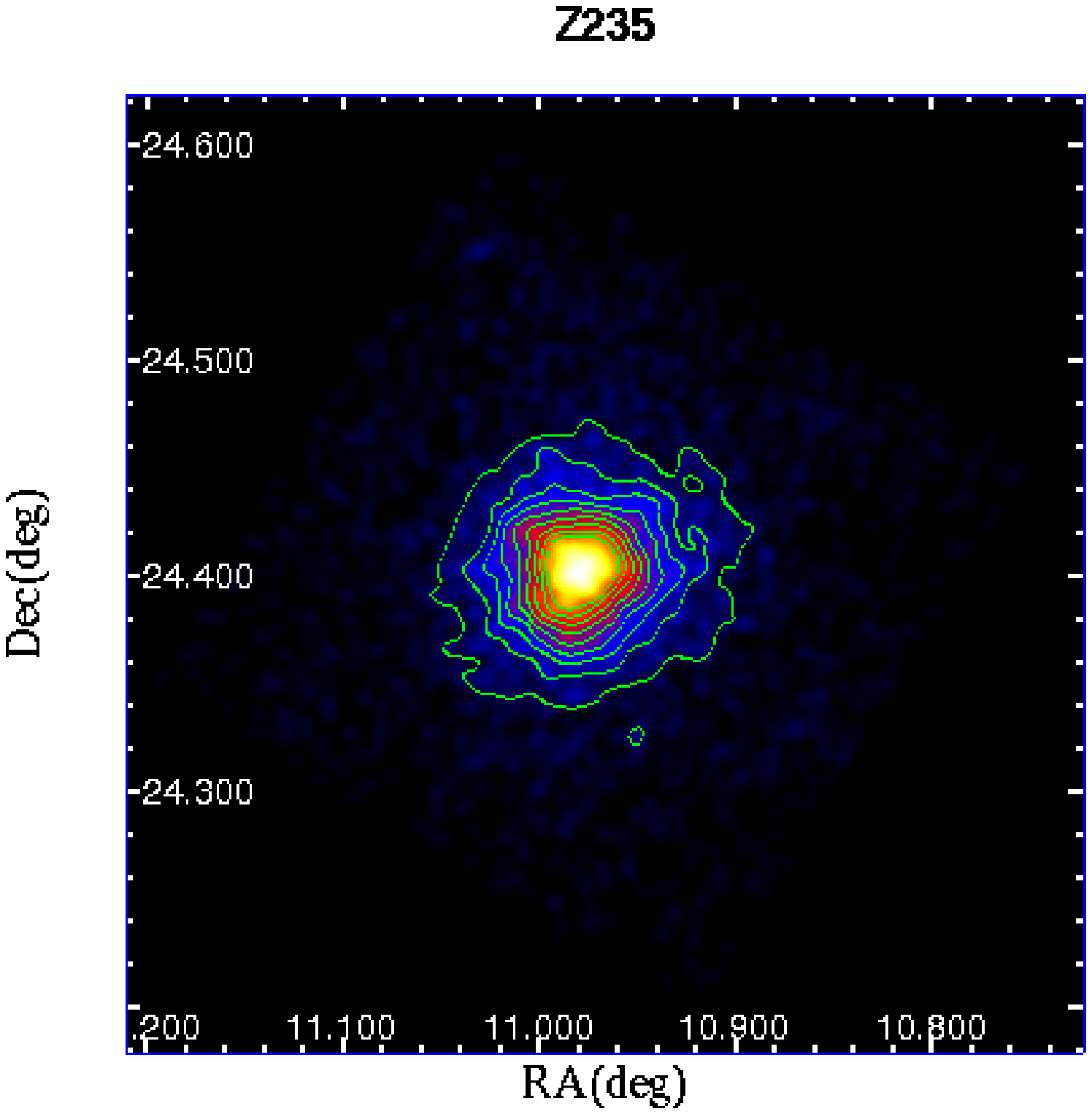}}}\\
\rotatebox{-0}{\resizebox{90mm}{!}{\includegraphics{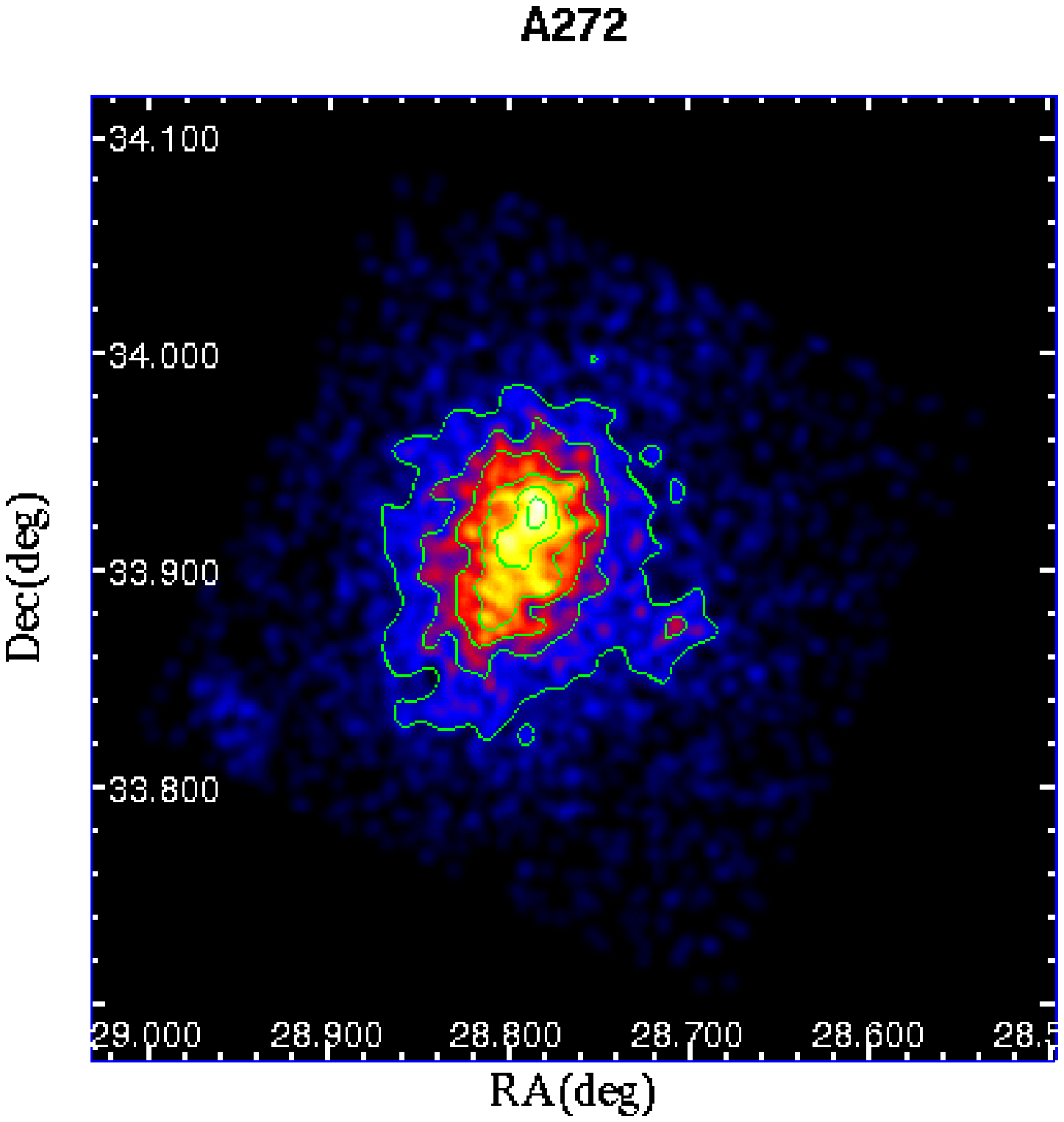}}}&
\rotatebox{-0}{\resizebox{90mm}{!}{\includegraphics{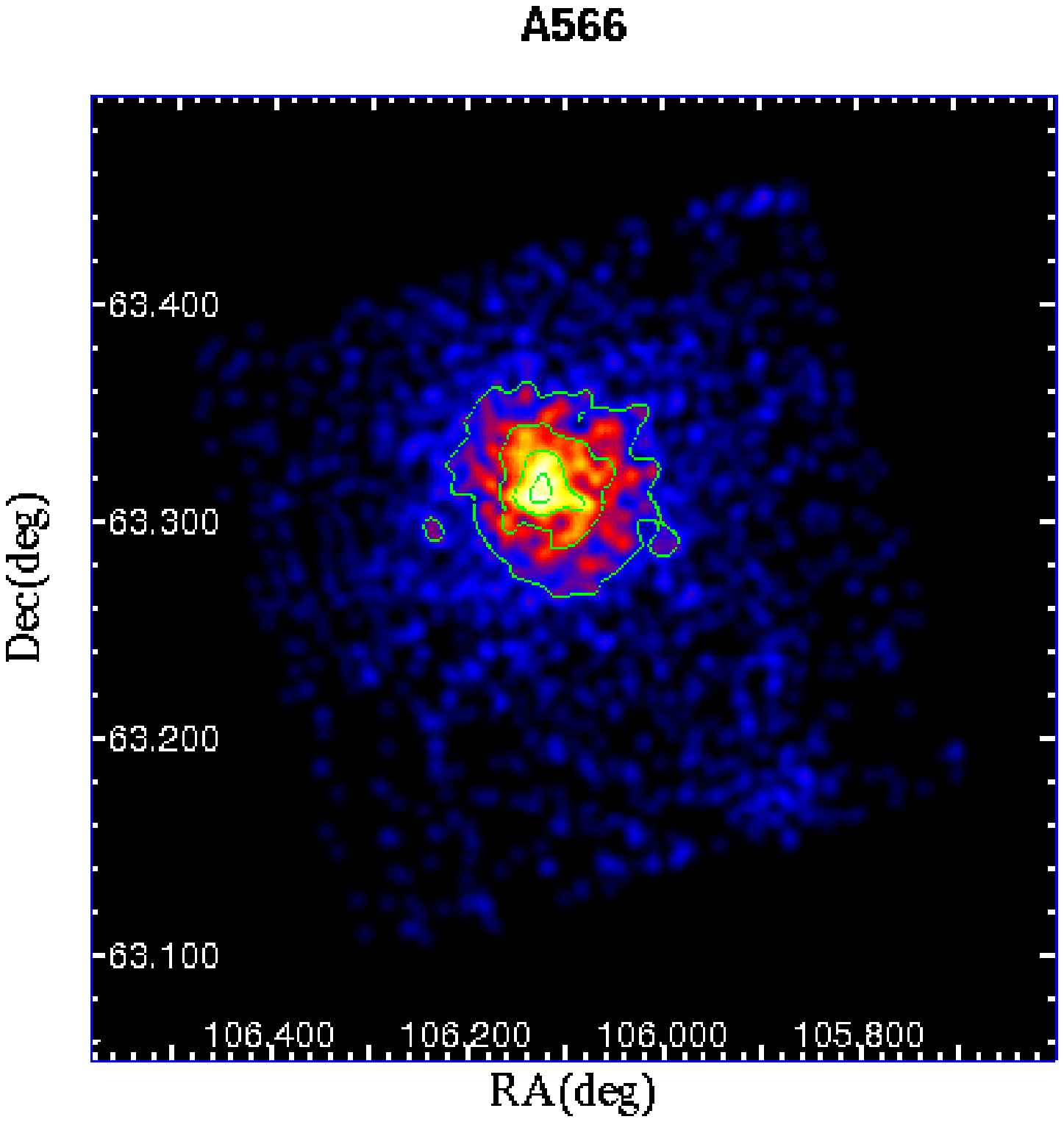}}}\\
\end{tabular}
\parbox{0.5\textwidth}{
\begin{minipage}{18cm}
\small\parindent=3.5mm
{\sc Fig.} \ref{fig:img}.--- Continued
\end{minipage}
}
%\caption{Continued}
%\label{fig:img}
\end{figure*}

\begin{figure}[tbg]
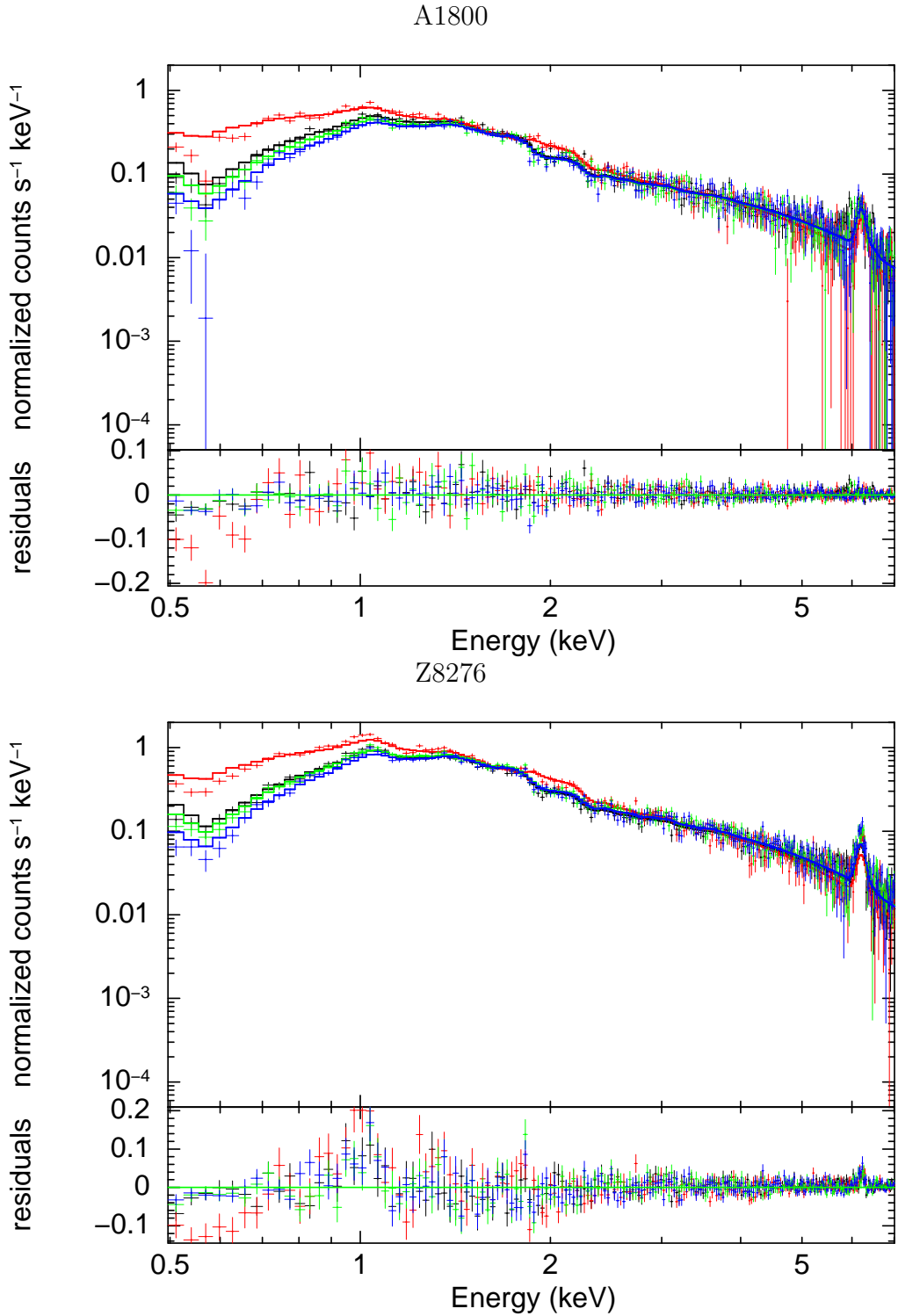

\epsscale{0.1}
\begin{tabular}{c}
A1800\\
\rotatebox{-90}{\resizebox{90mm}{!}{\includegraphics{f3a.ps}}}\\
%\rotatebox{-90}{\resizebox{50mm}{!}{\includegraphics{A1800_2T.ps}}}\\
Z8276\\
\rotatebox{-90}{\resizebox{90mm}{!}{\includegraphics{f3b.ps}}}\\
%\rotatebox{-90}{\resizebox{50mm}{!}{\includegraphics{Z8276_2T.ps}}}

\end{tabular}

\caption{Spectra of A1800 and Z8276 and 1T model fit in the energy range
0.5-7 keV, background has already been subtracted. According to our
F-test, A1800 is a NCCC, while Z8276 is CCC. Spectra from XIS0, XIS1,
XIS2 and XIS3 are shown in black, red, green and blue, respectively.}
\label{fig:spec}
\end{figure}

\begin{figure}[tbg]
\epsscale{0.1}
\begin{tabular}{c}

\rotatebox{-0}{\resizebox{150mm}{!}{\includegraphics{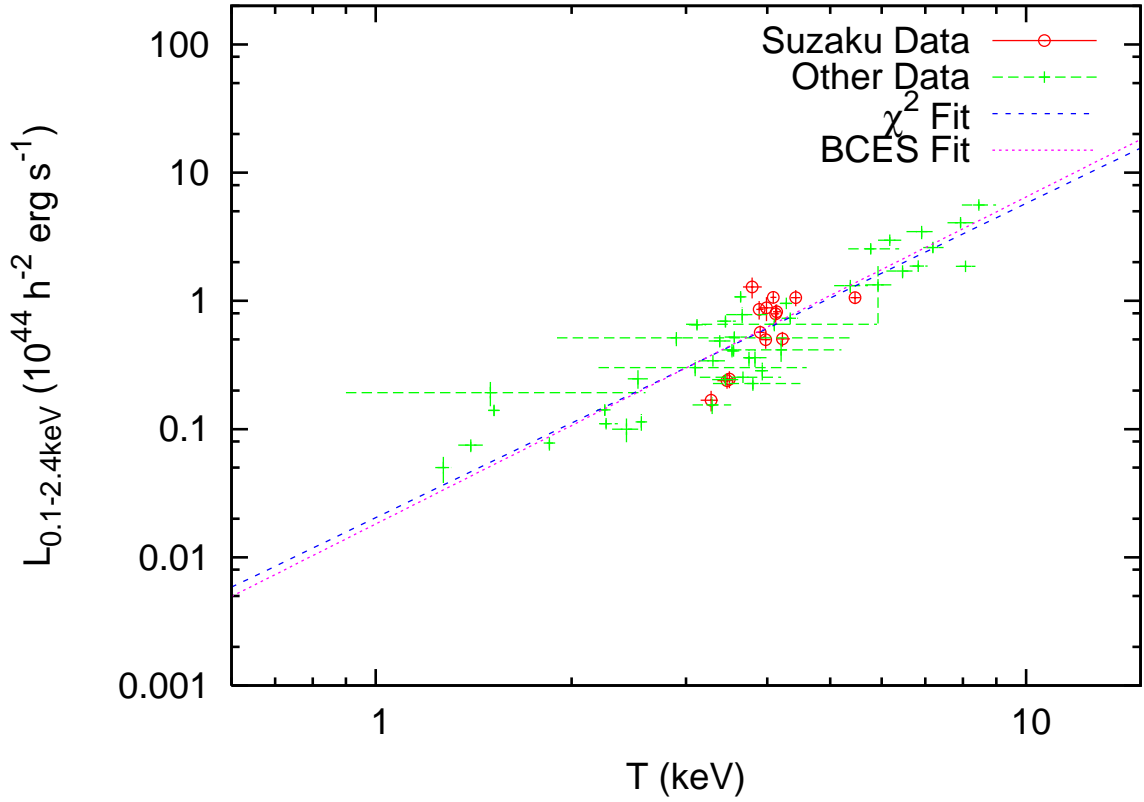}}}
\end{tabular}
\caption{Cluster
luminosity in the energy range of 0.1-2.4 keV is plotted versus 
temperature. The new {\sl Suzaku} data is shown together with the data from the rest of our flux-limited sample. Both the BCES best fit $L-T$ relation and that of our modified $\chi^2$ method are shown.
\label{fig:data}}
\end{figure}

\begin{figure}[tbg]
\epsscale{0.1}
\begin{tabular}{c}

\rotatebox{-0}{\resizebox{150mm}{!}{\includegraphics{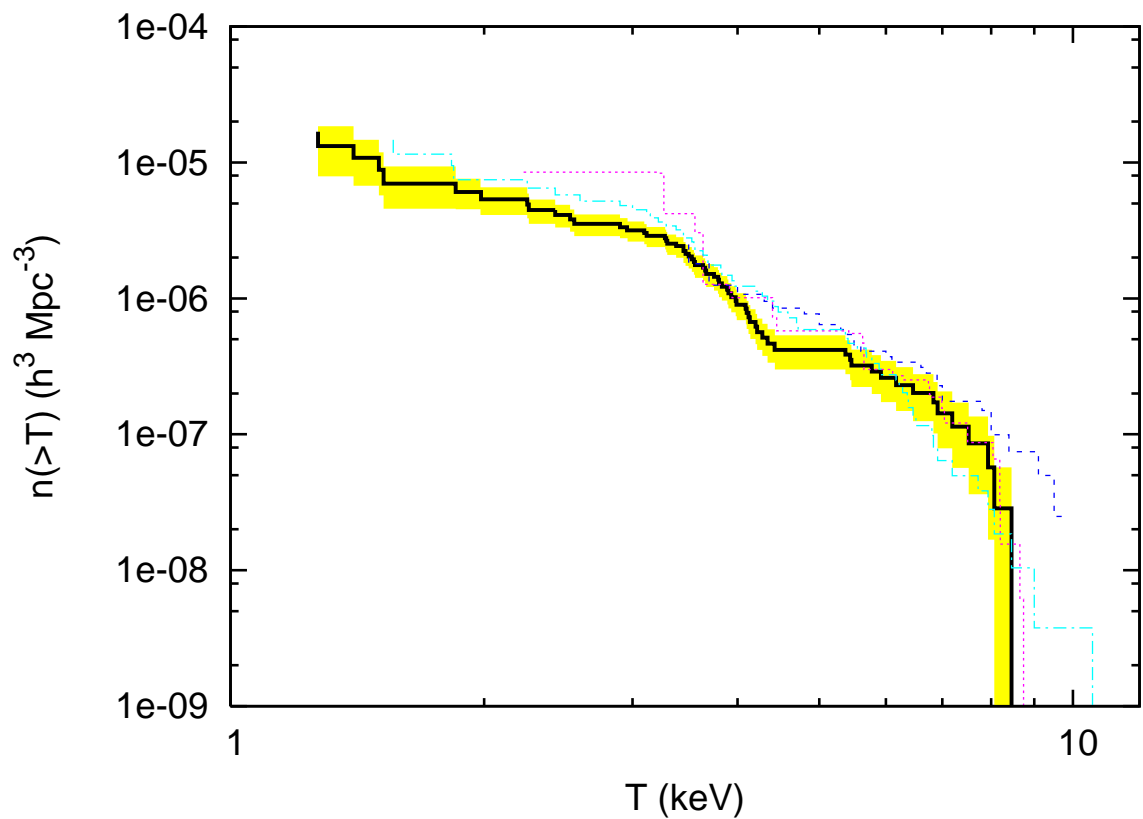}}}
\end{tabular}
\caption{
Cumulative XTF together with previous results. The thick black solid line
is our present result, the red dotted line is from I02, the blue short-dashed 
line is from Markevitch (1998), the cyan dash-dotted line is from Henry (2000). The shaded (yellow)
region indicates the 1-$\sigma$ poisson error on our XTF measurement.}
\label{fig:xtf}
\end{figure}

\begin{figure}[tbg]
\epsscale{0.1}
\begin{tabular}{c}

\rotatebox{-0}{\resizebox{150mm}{!}{\includegraphics{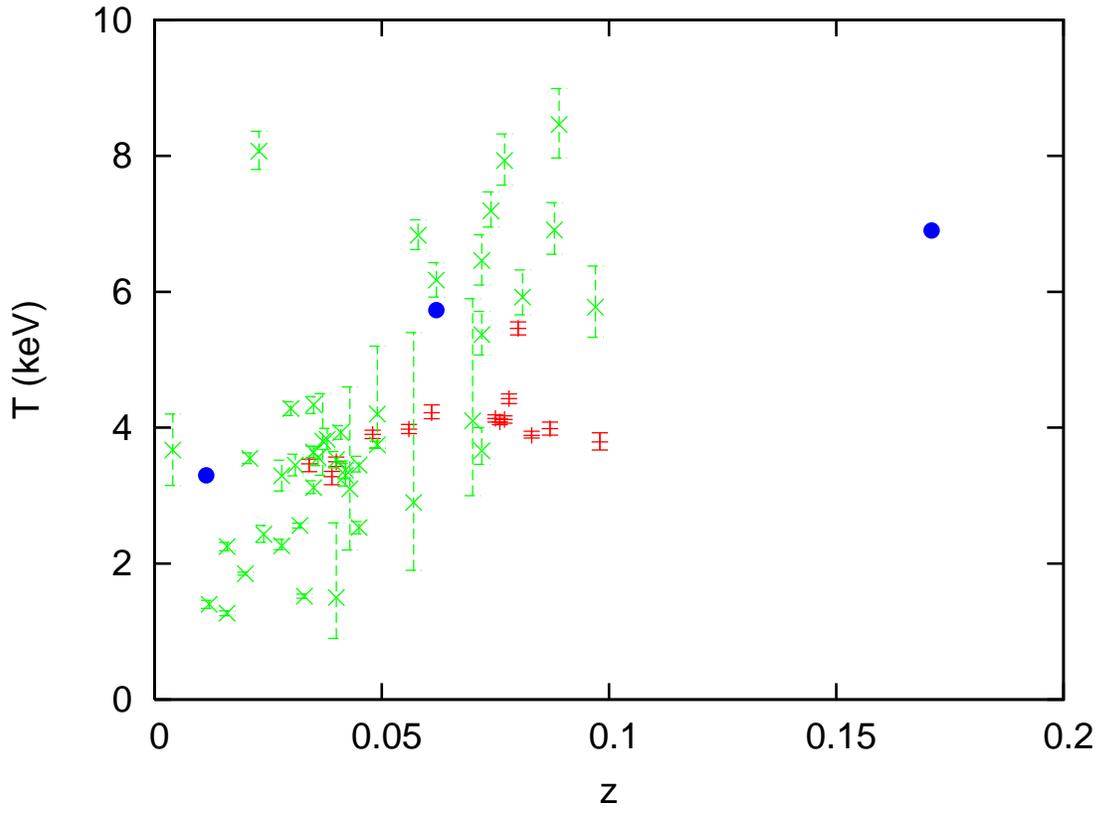}}}
\end{tabular}
\caption{Cluster temperature is plotted versus redshift for our sample. {\sl Suzaku} cluster measurements are shown in red. The three objects used to compare our X-ray temperature analysis to that of earlier missions (Table 4) are plotted as filled circles.
}
\label{fig:t_z}
\end{figure}
 
 \clearpage

\begin{table*}[ht]
  \caption{Suzaku observations of 14 galaxy clusters }
    \label{tbl:suzaku}
\begin{center}
\begin{tabular}{ c c c c c}
\tableline \tableline
Cluster Name & Sequence Number & (RA,DEC)(deg) & Exposure Time (ks) &  Processing Version\\
\tableline
UGC03957 & 801072010 & (115.2386, 55.4389)& 12.4 & 1.2.2.3\\
A2572b & 801073010 & (349.6122, 18.7237) & 22.4 & 1.2.2.3\\
Z8852 & 801074010 & (347.6298, 7.5832)& 8.6 & 1.2.2.5\\
II Zw 108 & 801075010 & (318.5265, 2.5682) & 11.4 & 1.2.2.3\\
A2665 & 801076010 & (357.7129, 6.1379)& 10.5 & 1.2.2.5\\
A1831 & 801077010 & (209.8017, 27.9785)& 8.9 & 1.2.2.3\\
A1800 & 801078010 & (207.3647, 28.1066) & 18.5 & 1.2.2.3\\
Z8276 & 801079010 & (266.0553, 32.9813) & 14.9 & 1.2.2.3\\
A2495 & 801080020 & (342.5710, 10.9166) & 11.9 & 1.4.2.9\\
A2061 & 801081010 & (230.3133, 30.6555) & 15.3 & 1.2.2.3\\
A2249 & 801082010 & (257.4518, 34.4734) & 22.8 & 1.3.2.6\\
Z235 & 801083010 & (10.9674, 24.4018) & 12.9 & 1.2.2.3\\
A272 & 801084010 & (28.7653, 33.8978) & 14.1 & 1.2.2.3\\
A566 & 801085010 & (106.0943, 63.2750) & 8.79 & 1.3.2.8\\
 \tableline 
\end{tabular}
\end{center}
\end{table*}

\clearpage

%\label{sec:data}
\begin{table*}[ht]
  \caption{Spectral Fitting Result (Metallicity fixed)}
    \label{tbl:temperature}
\begin{center} \small
\begin{tabular}{ l c c c r l l r}
\tableline \tableline
Cluster Name & Redshift& $n_{\mathrm{H}}$ & \multicolumn{2}{c}{1T Model} & \multicolumn{3}{c}{2T Model} \\
& &($10^{20}\mbox{cm}^{-2})$&$kT$(keV)&$\chi^{2}/d.o.f$&$kT_{h}$(keV)&$kT_{l}$(keV)&$\chi^{2}/d.o.f$\\
\tableline
UGC03957 & 0.0341 & 4.35 & $2.49_{-0.04}^{+0.04}$ &1495.19/887&  ${\bf 3.47_{-0.19}^{+0.11}}$ & $1.84_{-0.03}^{+0.07}$ &1457.29/885\\
A2572b & 0.0389 & 4.10 & $2.73_{-0.04}^{+0.04}$ & 1566.49/887&${\bf 3.28_{-0.19}^{+0.13}}$ & $1.24_{-0.12}^{+0.08}$ &1424.93/885\\
Z8852 & 0.0400 & 4.73 & ${\bf 3.50_{-0.11}^{+0.11}}$ & 1013.11/665&$3.70_{-0.96}^{+0.78}$ & $3.17_{-0.82}^{+1.64}$ &1013.12/663\\
II Zw 108 & 0.0483 & 5.85& ${\bf 3.90_{-0.10}^{+0.10}}$ & 1043.86/887&$3.97_{-0.11}^{+0.11}$& $1.13_{-0.61}^{+4.08}$ &1041.17/885\\
A2665 & 0.0562 & 5.82 & ${\bf 3.98_{-0.11}^{+0.11}}$ & 1038.76/665&$14.31_{-4.40}^{+9.97}$ & $2.98_{-0.14}^{+0.21}$ &1015.34/663\\
A1831 & 0.0612 & 1.37 & $3.65_{-0.06}^{+0.07}$ & 1286.31/887&${\bf 4.22_{-0.14}^{+0.19}}$ & $1.14_{-0.06}^{+0.15}$ &1178.48/885\\
A1800 & 0.0748 & 1.17 & ${\bf 4.14_{-0.09}^{+0.09}}$ & 1196.85/887&$4.19_{-0.46}^{+0.33}$ &$3.87_{-1.50}^{+2.86}$&1196.85/885\\
Z8276 & 0.0757 & 3.88 & $3.58_{-0.05}^{+0.05}$ & 1252.13/887&${\bf 4.09_{-0.07}^{+0.08}}$ &$1.45_{-0.03}^{+0.08}$&1176.05/885\\
A2495 & 0.0768 & 4.82 & ${\bf 4.12_{-0.09}^{+0.09}}$ & 1243.78/665&$8.71_{-2.27}^{+4.33}$ & $3.71_{-0.13}^{+0.55}$&1241.32/663\\
A2061 & 0.0777 & 1.67 & $4.30_{-0.10}^{+0.10}$ & 1035.92/887&${\bf 4.43_{-0.12}^{+0.12}}$ & $0.71_{-0.14}^{+0.14}$&1024.50/885\\
A2249 & 0.0802 & 2.22 & ${\bf 5.46_{-0.16}^{+0.16}}$ & 879.93/665&$12.10_{-4.57}^{+6.95}$ & $4.62_{-0.22}^{+0.42}$&875.49/663\\
Z235 & 0.0830 & 3.73 & $3.52_{-0.06}^{+0.06}$ & 1451.43/887&${\bf 3.89_{-0.06}^{+0.10}}$ & $1.15_{-0.02}^{+0.07}$&1343.01/885\\
A272 & 0.0872 & 4.99 & $3.26_{-0.07}^{+0.07}$ & 1125.81/887&${\bf 3.99_{-0.16}^{+0.16}}$ & $1.71_{-0.06}^{+0.10}$&1104.59/885\\
A566 &  0.0980 & 5.03 & ${\bf 3.79_{-0.12}^{+0.13}}$ & 1098.36/665&$43.47_{}^{}$&$3.86_{-0.10}^{+0.14}$&1096.52/663 \\
 \tableline 
%\tablenotetext{}
\end{tabular}
\tablenotetext{}{{\sc Notes}--- Errors in the tables are all 90\% error.}
\end{center}
\end{table*}

\clearpage

\begin{table*}[ht]
  \caption{Spectral Fitting Result (Metallicity as free parameter)}
    \label{tbl:metallicity}
\begin{center} \small
\begin{tabular}{ l c c c c c c c}
\tableline \tableline
Cluster Name & \multicolumn{3}{c}{1T Model} & \multicolumn{4}{c}{2T Model} \\
& $kT$(keV)&$Z$($Z_{\odot}$)&$\chi^{2}/d.o.f$&$kT_{h}$(keV)&$kT_{l}$(keV) &$Z$($Z_{\odot}$)&$\chi^{2}/d.o.f$\\
\tableline
UGC03957 & $\bf {2.60_{-0.04}^{+0.04}}$ &${\bf 0.59_{-0.03}^{+0.03}}$&1228.23/886&  $2.64_{-0.16}^{+0.31}$&$2.55_{-0.35}^{+0.19}$ &$0.59_{-0.03}^{+0.03}$ &1228.24/884\\
A2572b & $2.72_{-0.05}^{+0.05}$ &$0.26_{-0.02}^{+0.02}$& 1560.63/886&${\bf 3.54_{-0.26}^{+0.34}}$&$1.35_{-0.12}^{+0.07}$ & ${\bf 0.23_{-0.03}^{+0.02}}$& 1409.27/884\\
Z8852 & ${\bf 3.50_{-0.11}^{+0.11}}$ & ${\bf0.32_{-0.05}^{+0.06}}$&1012.76/664&$3.69_{-0.82}^{+0.82}$&$3.18_{-0.82}^{+1.65}$&$0.32_{-0.05}^{+0.06}$ &1012.77/662\\
II Zw 108 & ${\bf 3.93_{-0.10}^{+0.10}}$ &${\bf 0.45_{-0.04}^{+0.04}}$& 1009.83/886&$4.03_{-0.10}^{+0.10}$&$0.68_{-0.16}^{+0.22}$ &$0.46_{-0.04}^{+0.05}$&1003.58/884\\
A2665 & ${\bf 3.96_{-0.11}^{+0.11}}$ &${\bf 0.35_{-0.04}^{+0.04}}$& 1035.00/664&$18.30_{-6.38}^{+13.10}$&$3.07_{-0.15}^{+0.17}$ &$0.41_{-0.05}^{+0.04}$&1000.71/662\\
A1831 & $3.64_{-0.06}^{+0.06}$ &$0.31_{-0.03}^{+0.03}$& 1270.34/886&${\bf 4.19_{-0.15}^{+0.20}}$&$1.14_{-0.07}^{+0.15}$ &${\bf 0.31_{-0.03}^{+0.03}}$&1161.77/884\\
A1800 & ${\bf 4.13_{-0.09}^{+0.09}}$ & ${\bf 0.36_{-0.03}^{+0.03}}$&1189.41/886&$4.19_{-0.82}^{+0.82}$&$3.87_{-0.82}^{+3.29}$ &$0.36_{-0.03}^{+0.03}$&1189.43/884\\
Z8276 & $3.59_{-0.04}^{+0.02}$ &$0.50_{-0.03}^{+0.03}$& 1109.93/886&${\bf 4.08_{-0.09}^{+0.09}}$&$1.71_{-0.10}^{+0.09}$ &${\bf 0.46_{-0.03}^{+0.03}}$&1081.53/884\\
A2495 & ${\bf 4.10_{-0.09}^{+0.09}}$ &${\bf 0.35_{-0.03}^{+0.04}}$& 1237.65/664&$11.98_{-4.11}^{+4.50}$&$3.60_{-0.09}^{+0.13}$ &$0.38_{-0.04}^{+0.04}$&1229.03/662\\
A2061 & $4.31_{-0.10}^{+0.11}$ &$0.24_{-0.03}^{+0.03}$& 1027.97/886&${\bf 4.44_{-0.13}^{+0.12}}$&$0.71_{-0.15}^{+0.19}$ &${\bf 0.25_{-0.03}^{+0.04}}$&1018.67/884\\
A2249 & ${\bf 5.52_{-0.16}^{+0.16}}$ &${\bf 0.20_{-0.04}^{+0.04}}$& 862.40/664&$5.84_{-1.64}^{+6.58}$&$5.43_{-0.82}^{+0.82}$ &$0.20_{-0.04}^{+0.04}$&862.42/662\\
Z235 & $3.51_{-0.05}^{+0.05}$ &$0.45_{-0.03}^{+0.03}$& 1368.40/886&${\bf 3.80_{-0.08}^{+0.07}}$&$1.14_{-0.05}^{+0.05}$ &${\bf 0.46_{-0.04}^{+0.02}}$&1292.94/884\\
A272 & $3.26_{-0.07}^{+0.07}$ &$0.31_{-0.03}^{+0.03}$& 1125.61/886&${\bf 4.03_{-0.18}^{+0.16}}$&$1.67_{-0.13}^{+0.11}$ &${\bf 0.26_{-0.03}^{+0.03}}$&1101.16/884\\
A566 &${\bf 3.79_{-0.12}^{+0.13}}$ &${\bf 0.33_{-0.04}^{+0.05}}$& 1096.84/664&$31.16_{}^{}$&$3.67_{-0.10}^{+0.13}$&$0.34_{-0.04}^{+0.05}$&1094.36/662 \\
 \tableline 
\end{tabular}
\end{center}
\end{table*}

\clearpage

%another table for consistent check
\begin{table}[ht]
  \caption{Calibration clusters}
    \label{tbl:observation} 
\begin{tabular}{c c c c}
\tableline \tableline
Cluster Name & $T_{\mbox{new}}$(keV) & $T_{\mbox{old}}$(keV) & Reference\\
\tableline
A1060 & 3.30 & 3.1-3.9 & 3,4,6,7,8,13,14,15\\
A2218 & 6.90 & 6.7-7.2 & 1,3,5,9,11,12\\
A1795 & 5.73 & 5.3-7.8 &2,3,8,10,16,17\\
 \tableline 
\end{tabular}
\tablenotetext{}{{\sc Reference}---(1)Allen 1998; (2)Allen et al. 2001;(3)David et al. 1993;
(4)Furusho et al. 2001; (5)Govoni et al. 2004; (6)Hayakawa et al. 2004; (7)Hayakawa et al. 2006; 
(8)I02;(9)Machacek et al. 2002; (10)Markevitch 1998; (11)McHardy et al. 1990; (12)Mushotzky \& Loewenstein 1997; 
(13)Sato et al. 2007;(14)Tamura et al. 1996; (15)Tamura et al. 2000; (16)Tamura et al. 2001;
(17)Vikhlinin et al. 2005}
\end{table}

\clearpage

%\label{sec:data}
\begin{table*}[ht]
  \caption{Complete Local Flux Limited Cluster Sample}
    \label{tbl:sample}
\begin{center}\scriptsize
\begin{tabular}{ l c c c c c c c}
\tableline \tableline
Cluster Name & Redshift& $kT$\tablenotemark{\alpha} & Flux & Luminosity&$V_{max}$&Instrument&Reference \\
 & &(keV)&($10^{-12} \mbox{erg cm}^{-2} \mbox{ s}^{-1}$)&($10^{44}h^{-2}\mbox{erg}\mbox{ s}^{-1}$)&($h^{-3}\mbox{Mpc}^{-3}$)&&\\
 \tableline
Virgo & 0.0036 & $3.67_{-0.52}^{+0.53}$ &$1821.1\pm 38.8$&$0.254\pm 0.005 $& 1.23E+07&ASCA& 11\\
RXJ0419.6+0225& 0.0123 &$1.4_{-0.06}^{+0.06}$&$45.7\pm 5.4$&$0.075\pm 0.009$& 4.03E+05 &ASCA& 8\\
A262&0.0163 &$2.25_{-0.06}^{+0.06}$&$48.6\pm 4.8$&$0.141\pm 0.014$& 2.20E+06 &ASCA& 7\\
RXJ0123.6+3315& 0.0164&$1.27_{-0.04}^{+0.04}$&$17.1\pm 4.0$&$0.050\pm 0.012$& 2.82E+05 &ASCA& 1\\
Z4803& 0.0200&$1.85_{-0.02}^{+0.02}$&$17.7\pm 2.2$&$0.078\pm 0.010$& 1.10E+06 &ASCA& 5\\
A1367& 0.0214&$3.55_{-0.08}^{+0.08}$&$81.4\pm 4.9$&$0.408\pm 0.024$& 1.10E+07 &ASCA& 7\\
RXJ0751.3+5012& 0.0220&$1.98^{e}$&$10.1\pm 1.8$&$0.054\pm0.010$& 1.40E+06 & \\
A1656& 0.0231&$8.07_{-0.27}^{+0.29}$&$319.2\pm 8.3$&$1.863\pm 0.048$& 3.50E+07 &ASCA& 7\\
A400& 0.0238&$2.43_{-0.12}^{+0.13}$&$16.2\pm 3.3$&$0.100\pm 0.021$& 2.90E+06 &ASCA& 7\\
MKW8& 0.0276&$3.29_{-0.22}^{+0.23}$&$18.4\pm 2.7$&$0.154\pm 0.023$& 8.50E+06 &ASCA& 7\\
RXJ1715.3+5725& 0.0280&$2.26_{-0.06}^{+0.10}$&$12.7\pm 0.9$&$0.110\pm0.008$& 2.24E+06 &ASCA& 6\\
RXJ0341.3+1524& 0.0290&$2.96^{e}$&$15.5\pm 3.1$&$0.142\pm0.028$& 5.83E+06 & \\
A2199& 0.0299&$4.28_{-0.10}^{+0.10}$&$96.8\pm 3.5$&$0.955\pm 0.034$& 1.97E+07 &ASCA& 7\\
A2634& 0.0309&$3.45_{-0.16}^{+0.16}$&$23.1\pm 2.5$&$0.243\pm 0.026$& 1.00E+07 &ASCA& 7\\
AWM4& 0.0318&$2.56_{-0.04}^{+0.04}$&$10.2\pm 1.5$&$0.114\pm 0.017$& 3.49E+06 &XMM& 10\\
RXJ1733.0+4345& 0.0330&$1.52_{-0.03}^{+0.03}$&$11.6\pm 1.1$&$0.140\pm0.013$& 5.42E+05 &Chandra & 4\\
UGC03957& 0.0341&$3.47_{-0.12}^{+0.07}$&$18.6\pm 2.6$&$0.239\pm 0.033$& 1.02E+07 &Suzaku& \\
2A0335+096& 0.0349&$3.64_{-0.08}^{+0.09}$&$80.5\pm 6.5$&$1.073\pm 0.087$& 1.20E+07 &ASCA& 7\\
A2147& 0.0353&$4.34_{-0.13}^{+0.12}$&$53.2\pm 3.7$&$0.732\pm 0.051$& 2.04E+07 &ASCA& 7\\
A2052& 0.0353&$3.12_{-0.09}^{+0.10}$&$47.1\pm 3.7$&$0.650\pm 0.051$& 7.06E+06 &ASCA& 7\\
A2063& 0.0355&$3.56_{-0.12}^{+0.16}$&$37.5\pm 3.4$&$0.523\pm 0.048$& 1.12E+07 &ASCA& 7\\
A2151a& 0.0370&$3.80_{-0.50}^{+0.70}$&$15.0\pm 1.9$&$0.227\pm 0.028$& 1.38E+07 &Einstein& 2\\
A576& 0.0381&$3.83_{-0.15}^{+0.16}$&$22.4\pm 3.5$&$0.360\pm 0.056$& 1.42E+07 &ASCA& 7\\
A2572b& 0.0389&$3.28_{-0.12}^{+0.08}$&$10.0\pm 1.9$&$0.168\pm 0.031$& 8.40E+06 &Suzaku&  \\
A76& 0.0395&$1.50_{-0.60}^{+1.10}$&$11.0\pm 2.4$&$0.192\pm 0.041$& 5.17E+05 &Einstein& 2\\
A2657& 0.0400&$3.53_{-0.12}^{+0.12}$&$23.4\pm 2.9$&$0.415\pm 0.052$& 1.08E+07 &ASCA& 7\\
Z8852& 0.0400&$3.50_{-0.07}^{+0.07}$&$13.8\pm 2.1$&$0.245\pm 0.037$& 1.05E+07 &Suzaku&  \\
A2107& 0.0411&$3.93_{-0.10}^{+0.10}$&$15.2\pm 2.4$&$0.285\pm 0.045$& 1.54E+07 &ASCA& 9\\
A2589& 0.0416&$3.38_{-0.13}^{+0.13}$&$25.3\pm 2.8$&$0.487\pm 0.055$& 9.34E+06 &ASCA& 7\\
HCG94& 0.0422&$3.30_{-0.16}^{+0.17}$&$17.2\pm 0.03$&$0.341\pm 0.006$& 8.60E+06 &ASCA& 7\\
A2593& 0.0428&$3.10_{-0.90}^{+1.50}$&$14.8\pm 2.1$&$0.302\pm 0.044$& 6.90E+06 &Einstein& 2\\
A168& 0.0448&$2.53_{-0.09}^{+0.09}$&$11.0\pm 1.8$&$0.247\pm 0.041$& 3.35E+06 &Chandra & 12\\
MKW3s& 0.0453&$3.45_{-0.10}^{+0.13}$&$30.3\pm 3.3$&$0.694\pm 0.075$& 1.00E+07 &ASCA& 7\\
IIZw108& 0.0483&$3.90_{-0.06}^{+0.06}$&$21.9\pm 2.8$&$0.569\pm 0.073$& 1.50E+07 &Suzaku&   \\
A376& 0.0488&$3.75_{-0.05}^{+0.05}$&$13.6\pm 2.0$&$0.360\pm 0.053$& 1.32E+07 &ASCA& 3    \\
A193& 0.0491&$4.20_{-0.50}^{+1.00}$&$15.4\pm 3.0$&$0.414\pm 0.080$& 1.87E+07 &EXOSAT& 2\\
A2665& 0.0562&$3.98_{-0.07}^{+0.07}$&$14.1\pm 2.2$&$0.499\pm 0.076$& 1.60E+07 &Suzaku&   \\
\tableline 
\end{tabular}
\end{center}
\end{table*}

\clearpage

\begin{table*}[ht]
\begin{center}\scriptsize
\begin{tabular}{ l c c c c c c c}
\tableline \tableline
Cluster Name & Redshift& $kT$\tablenotemark{\alpha} & Flux & Luminosity&$V_{max}$&Instrument&Reference \\
 & &(keV)&($10^{-12} \mbox{erg cm}^{-2} \mbox{ s}^{-1}$)&($10^{44}h^{-2}\mbox{erg}\mbox{ s}^{-1}$)&($h^{-3}\mbox{Mpc}^{-3}$)&&\\
 \tableline
A2626& 0.0565&$2.90_{-1.00}^{+2.50}$&$14.3\pm 2.0$&$0.515\pm 0.071$& 5.45E+06 &Einstein& 2\\
A2256& 0.0581&$6.83_{-0.21}^{+0.23}$&$49.5\pm 2.2$&$1.867\pm 0.082$& 3.46E+07 &ASCA& 7\\
A1831& 0.0612&$4.22_{-0.09}^{+0.12}$&$11.9\pm 1.6$&$0.505\pm 0.067$& 1.89E+07 &Suzaku&   \\
A1795& 0.0622&$6.17_{-0.25}^{+0.26}$&$68.1\pm 4.0$&$2.976\pm 0.173$& 3.37E+07 &ASCA& 7\\
A1767& 0.0701&$4.10_{-1.10}^{+1.80}$&$11.7\pm 1.4$&$0.656\pm 0.076$ & 1.74E+07 &Einstein& 2\\
A1589& 0.0718&$5.44^{e}$&$10.8\pm 1.6$&$0.634\pm0.094$& 3.09E+07 & \\
A399& 0.0722&$6.46_{-0.36}^{+0.38}$&$28.8\pm 3.5$&$1.709\pm 0.208$& 3.42E+07 &ASCA& 7\\
A2065& 0.0723&$5.37_{-0.30}^{+0.34}$&$22.2\pm 2.8$&$1.317\pm 0.166$& 3.04E+07 &ASCA& 7\\
A1775& 0.0724&$3.66_{-0.20}^{+0.34}$&$13.0\pm 1.9$&$0.782\pm 0.112$& 1.22E+07 &ASCA& 7\\
A401& 0.0739&$7.19_{-0.24}^{0.28}$&$42.8\pm 3.9$&$2.607\pm 0.237$& 3.48E+07 &ASCA& 7\\
A1800& 0.0748&$4.14_{-0.06}^{0.06}$&$12.8\pm 1.8$&$0.822\pm 0.115$& 1.79E+07 &Suzaku&   \\
Z5029& 0.0750&$7.52^{e}$&$22.0\pm 2.4$&$1.404\pm0.155$& 3.49E+07 & \\
Z8276& 0.0757&$4.09_{-0.04}^{+0.05}$&$16.4\pm 1.7$&$1.604\pm 0.111$& 1.73E+07 &Suzaku&   \\
A2029& 0.0766&$7.93_{-0.04}^{+0.04}$&$61.3\pm 4.3$&$4.068\pm 0.288$& 3.50E+07 &ASCA& 7\\
A2495& 0.0768&$4.12_{-0.05}^{0.05}$&$11.8\pm 2.1$&$0.792\pm 0.141$& 1.77E+07 &Suzaku&   \\
A2061& 0.0777&$4.43_{-0.07}^{+0.07}$&$15.3\pm 2.2$&$1.059\pm 0.153$& 2.14E+07 &Suzaku&\\
A2249& 0.0802&$5.46_{-0.09}^{+0.10}$&$14.4\pm 1.6$&$1.059\pm 0.115$& 3.10E+07 &Suzaku&\\
A2255& 0.0809&$5.92_{-0.26}^{0.40}$&$17.8\pm 8.9$&$1.330\pm 0.665$& 3.30E+07 &ASCA& 7\\
Z235& 0.0830&$4.31_{-0.06}^{+0.06}$&$10.9\pm 1.8$&$0.856\pm 0.140$& 2.00E+07 &Suzaku& \\
A272& 0.0872&$3.99_{-0.10}^{+0.10}$&$10.1\pm 2.1$&$0.881\pm 0.186$& 1.61E+07 &Suzaku&   \\
A478& 0.0882&$6.91_{-0.36}^{+0.40}$&$39.9\pm 5.1$&$3.465\pm 0.441$& 3.47E+07 &ASCA& 7\\
A2142& 0.0894&$8.46_{-0.49}^{+0.53}$&$61.5\pm 3.9$&$5.589\pm 0.357$& 3.50E+07 &ASCA& 7\\
A2244& 0.0970&$5.77_{-0.44}^{+0.61}$&$23.4\pm 2.0$&$2.542\pm 0.215$& 3.25E+07 &ASCA& 7\\
A566& 0.0980&$3.79_{-0.12}^{+0.13}$&$11.6\pm 2.2$&$1.284\pm 0.239$& 1.35E+07 &Suzaku& \\
\tableline 
\end{tabular}
\tablenotetext{}{{\sc Notes}--- All the errors are 1 $\sigma$ error. The 90\% temperature error in literature is converted to 1 $\sigma$ error assuming it's Gaussian error.}
\tablenotetext{\alpha}{Temperature with superscript $e$ is temperature estimated from our $L-T$ relation.}
\tablenotetext{}{{\sc Reference}--- (1)Buote \& Fabian 1998 ; (2)David et al. 1993; (3)Dupke \& Bregman 2005; 
(4)Fukazawa et al. 2006;(5) Horner 2001;
(6)Hwang et al. 1999; (7)I02;
(8)Kawaharada et al. 2003; (9)Novicki et al. 2002, (10)O'Sullivan et al. 2005; (11)Shibata et al. 2001;
(12)Yang et al. 2004. 
}
\end{center}
\end{table*}

\clearpage

\begin{table*}[ht]
  \caption{$L-T$ Relation Fitting Results}
    \label{tbl:fit}
\begin{center}
\begin{tabular}{c c c c c c c}
\tableline \tableline

Fitting Method & \multicolumn{3}{c}{$L_{0.1-2.4}-T$}&\multicolumn{3}{c}{$L_{\mathrm{Bol}}-T$}\\
 & $\alpha$ & $\beta$ & $\sigma^{2}_{\mbox{intr}}$& $\alpha$ & $\beta$ & $\sigma^{2}_{\mbox{intr}}$\\
\tableline
$\chi^{2}$&42.31&2.45&0.18&42.31&2.95&0.18\\
BCES&$42.26\pm 0.12$&$2.55\pm 0.19$&0.18&$42.20\pm 0.14$&$3.16\pm 0.22$&0.18\\
 \tableline 
\end{tabular}
\end{center}
\end{table*}

\end{document}